\documentclass[tightenlines,showpacs,prd,floatfix,preprintnumbers,amsmath,amssymb,nofootinbib,superscriptaddress]{revtex4-1}
\DeclareMathAlphabet{\mathpzc}{OT1}{pzc}{m}{it}
\usepackage{graphicx}% Include figure filesg
\usepackage{amsfonts}
\usepackage{hyperref}
\usepackage[normalem]{ulem}
\usepackage{color}
\usepackage{mathrsfs}
\usepackage{bm}
\usepackage{comment}
\usepackage{ulem}

\def\up#1{\raise1mm\hbox{$\!\!^{#1}$}} 
\def\upp#1{\raise2mm\hbox{$\!\!\!\!^{#1}$}}

\newcommand{\beq}{\begin{equation}}
\newcommand{\be}{\begin{equation}}
\newcommand{\eeq}{\end{equation}}
\newcommand{\ee}{\end{equation}}
\newcommand{\bea}{\begin{eqnarray}}
\newcommand{\eea}{\end{eqnarray}}
\newcommand{\bal}{\begin{align}}
\newcommand{\eal}{\end{align}}

\newcommand{\beaa}{\begin{eqnarray*}} 
\newcommand{\eeaa}{\end{eqnarray*}}

\newcommand{\ep}{\epsilon}
\newcommand{\ka}{\kappa}

\newcommand{\bsube}{\begin{subequations}}
\newcommand{\esube}{\end{subequations}}

\newcommand{\cst}{\cos\theta}
\newcommand{\snt}{\sin\theta}
\newcommand{\om}{\omega}
\newcommand{\tht}{\theta}

\begin{document}

%\title{Higher order post-Newtonian expansion of gravitational radiation flux for a particle in Kerr spacetime - I\\ (Circular, equatorial orbits)}

\title{Gravitational-wave flux for a particle orbiting a Kerr black hole to 20th post-Newtonian order: a numerical approach}

\author{Abhay G. Shah} 
\email{a.g.shah@soton.ac.uk}
\affiliation{School of Mathematics, University of Southampton, Southampton SO17 1BJ, United Kingdom}

%\pacs{04.30.Db, 04.25.Nx, 04.70.Bw}

\begin{abstract}
In this article we present the post-Newtonian (pN) coefficients of the energy flux (and angular momentum flux) at infinity and event horizon for a particle in circular, equatorial orbits about a Kerr black hole (of mass $M$ and spin-parameter $a$) up to 20-pN order. When a pN term is not a polynomial in $a/M$ and includes irrational functions (like polygamma functions), it is written as a power series of $a/M$. This is achieved by calculating the fluxes numerically with an accuracy greater than 1 part in $10^{600}$ using the methods outlined in \cite{SasTagLivRev, TMT}. Such high accuracy allows us to extract analytical values of pN coefficients that are linear combinations of transcendentals like the Euler constant, logarithms of prime numbers and powers of $\pi$. We also present the 22-pN expansion (spin-independent pN expansion) of the ingoing energy flux at the event horizon for a particle in circular orbit about a Schwarzschild black hole. 
%
%The pN expansion of the fluxes at infinity and horizon presented here are valid up to a few times the radius of the inner-most-stable-circular-orbit (ISCO) for a given $a/M$.the pN expansion of the individual $(\ell,m)$ modes, which are easily calculable from the numerics involved in this calculation, can be \emph{resummed} to provide a better convergence, even for radii lesser than the the ISCO radius for a given $a/M$.
\end{abstract}

\maketitle

\section{Introduction}
\label{sec1}
%What are EMRIs
Gravitational waves from extreme-mass-ratio inspirals (EMRIs) are an important target of the planned space-based gravitational wave observatory, \emph{eLISA}. In EMRIs a compact object (a neutron star or stellar-sized black hole) orbits a galactic, super-massive black hole, loses energy and angular momentum in form of gravitational waves emitted inwards and outwards, and spirals in. 
%self-force
Modeling EMRIs using numerical relativity is difficult because of vastly different length scales involved and they are best modeled using self-force formalism (or perturbation theory) which uses mass-ratio as its expansion parameter. To the lowest order (zeroth order) in mass-ratio the small compact object, now a test mass, follows a geodesic of the background. To first order in mass-ratio, the object perturbs the spacetime and the smooth part of this perturbation acts back on itself (\emph{self-force}) and alters its motion which will now be a geodesic of the background plus this smooth perturbation. 
%pN-theory
Another approach to model binaries is using the post-Newtonian theory. Far away from the black hole where the gravitational field is weak and the relative velocity is small compared to the speed of light, one can use post-Newtonian (pN) expansion of the Einstein equations to model its motion and calculate the gravitational waveforms. 
%JLF suggestion
In the pN approximation, an inspiraling quasi-circular orbit is determined by the orbital energy and the flux of gravitational waves. An overlapping regime where the pN and EMRI approximations are both valid allows one to use the EMRI approximation to find the pN expansion of the flux at first-order in the symmetric mass ratio $\frak{m}M/(M+\frak{m})^2$ and to arbitrarily high pN order. An EMRI computation of a gauge invariant quantity associated with the orbital redshift \cite{detweiler08,sbd08,sf3,sf4,bs11,Akcay2012ea} can similarly be used to find the coefficients in the orbital energy at first-order in the symmetric mass ratio and at arbitrarily high pN order \cite{BDTW11,sf5,BD2}.
% What am I doing there
%In this article we study the overlap region of the self-force formalism and pN-theory where we calculate the coefficients that appear in the pN-expansion of the energy and angular momentum luminosity to first order in mass-ratio. We use the gauge-invariant orbital frequency, $\Omega$, as our expansion parameter, i.e., the fluxes, to first order in mass-ratio, are expanded in negative powers of $R:=(M\Omega)^{-2/3}$.

%forget this - Recently, waveform models appropriate for data analysis were constructed for EMRIs in quasi-circular, equatorial orbits about a Kerr black hole using the effective-one-body (EOB) formalism in \cite{1009.6013}. Higher the pN order of fluxes known for EMRIs, greater the accuracy of the developed waveform for later stages in the inspiral.
 
%Horizon
Recently there has been in increasing interest in calculating the horizon-absorbed fluxes to study the tidal-heating and torquing of the Kerr black hole \cite{1211.1686} which affect the gravitational waves emitted by EMRIs by changing the intrinsic black hole parameters, and Taracchini et al. in \cite{1305.2184} proposed an improved model for the horizon-absorbed flux emitted by a particle in circular, equatorial orbit about a Kerr black hole based on earlier works (see references there in, \cite{BNZ12, NagarAkcay}). 
%nagar
%And the horizon-absorbed flux for the non-spinning case were studied by Bernuzzi et al. in \cite{BNZ}.
%
PN expansion of the absorbed fluxes for a particle in circular orbit about a Schwarzschild black hole was first performed in \cite{PoissonSasaki}; the study showed that the leading contribution enters at 4-pN order beyond the quadrupole formula. In \cite{TMT}, the flux absorbed by a Kerr black hole for a particle in circular, equatorial orbit was calculated  to 4pN beyond the lowest order and it was shown that in the case of rotating black holes the leading contribution enters at 2.5-pN order, i.e., earlier than in the case of non-rotating black holes. And recently R. Fujita calculated the flux to 8-pN order in the Kerr-case and about 17.5-pN order in the Schwarzschild-case \cite{FujitaKerr}.

% Infinity
Compared to the horizon-absorbed flux, a lot of work has been done on studying pN-expansion of fluxes at infinity. The first pN-expansion of flux at infinity for a particle in circular orbit about a Schwarzschild black hole was first performed in \cite{PoissonPN, CFPS, TagNak} to 1.5-pN order which was then extended to subsequently higher orders until recently when R. Fujita in \cite{Fujita} calculated it to a remarkable 22-pN order. For the case of a particle in circular, equatorial orbit about a Kerr black hole the energy luminosity at infinity was calculated in \cite{9603028} to 4pN order beyond the quadrupole formula, and recently R. Fujita calculated the flux to 10-pN order in \cite{FujitaKerr}.

% Our work 
Inspired by \cite{Fujita} and \cite{sf5} we, by working with an accuracy greater than 1 part in $10^{600}$ for orbits extending from about $10^{18}M$ to $10^{34}M$, calculate the pN coefficients of energy and angular momentum fluxes at (i) infinity and (ii) event horizon for a particle in circular equatorial orbit about a Kerr black hole to 20-pN order, and at (iii) the event horizon for a particle in circular orbit about a Schwarzschild black hole to 22-pN order. For relatively lower pN-order where the coefficients are simple, finite polynomial in $a/M$ ($a$ being the Kerr spin-parameter), with the help of \emph{Mathematica}, we are able to extract analytical coefficients which are not only rational numbers or rational numbers times $\pi$ but also linear combination of transcendentals like Euler's constant, logarithms of prime numbers, and powers of $\pi$. Since higher pN coefficients have irrational functions of $a/M$, like the polygamma functions of $a/M$ and $\sqrt{1-({a/M})^2}$, we present each pN coefficient as a power series in $a/M$ (at most pN orders we present the first 20 non-zero terms of this power series).

% Organization
This article is organized as follows. In Sect. \ref{sec2} we present the important equations involved in calculating the fluxes at infinity and event horizon. Since the calculations involved and derivations are very long, we refer the readers to \cite{SasTagLivRev, TMT, Hughes00} and the references therein for their derivation. In Sect. \ref{sec3} we present the analytical pN coefficients of energy fluxes for the three different cases mentioned above, and the numerical values of the remaining pN coefficients are presented in the accompanying text files.%Appendix \ref{appB}. 
We work in the $G=c=1$ units.

\section{Computation}
\label{sec2}

We consider a particle of mass $\mathfrak{m}$ orbiting a Kerr black hole of mass $M$ and spin parameter $a$ along direct, circular, equatorial orbits at radius $r=r_0$. At zeroth order in $\mathfrak{m}/M$, the trajectory is a circular orbit. In Kerr coordinates, its 4-velocity is 
given by
\begin{align} \allowdisplaybreaks
u^\alpha &= u^t (t^\alpha + \Omega \phi^\alpha),\,\textrm{where} \nonumber \\
u^t &= \frac{r_0^{3/2}+a M^{1/2}}{ \sqrt{ r_0^3-3Mr_0^2+2aM^{1/2}r_0^{3/2}} },\, \textrm{and} \nonumber  \\
\Omega &= \frac{M^{1/2}} { (r_0^{3/2} + a M^{1/2})},
\end{align}
where $\phi^\alpha$ and $t^\alpha$ are the rotational and asymptotically timelike Killing vectors of the Kerr spacetime, respectively. Its energy and angular momentum per unit mass are given by 
\bea
\hat{E} &=&  \frac{ r_0^{3/2} - 2M r_0^{1/2} + a M^{1/2} } { r_0^{3/4}(r_0^{3/2}-3M r_0^{1/2}+2aM^{1/2})^{1/2} } , \nonumber \\
\hat{L}_z &=& \frac{ M^{1/2}(r_0^2-2aM^{1/2}r_0^{1/2}+a^2) }{ r_0^{3/4}(r_0^{3/2}-3M r_0^{1/2}+2aM^{1/2})^{1/2} }.
\eea
In this section, we will not discuss the derivation of the equations involved in the calculations of energy flux and angular momentum flux, which are presented clearly in \cite{SasTagLivRev, TMT, Hughes00, Fujita}, and will only present the main equations required to calculate them. 
The energy and the angular momentum flux at infinity and event horizon for our case are given by 
\bea \label{lmmode}
\left\langle \frac{dE}{dt}\right\rangle_\infty &=& \sum_{\ell,m} \frac{\left| Z_{\ell,m}^\infty \right|^2}{4\pi \omega^2}, \, \nonumber \\
\left\langle \frac{dJ}{dt}\right\rangle_\infty &=& \frac{1}{\Omega}\left\langle \frac{dE}{dt}\right\rangle_\infty, \nonumber \\
\left\langle \frac{dE}{dt}\right\rangle_H &=& \sum_{\ell,m} \frac{\alpha_{\ell,m} \left| Z_{\ell,m}^H \right|^2}{4\pi \omega^2}, \, \textrm{and} \nonumber \\
\left\langle \frac{dJ}{dt}\right\rangle_H &=& \frac{1}{\Omega}\left\langle \frac{dE}{dt}\right\rangle_H,
\eea
where $\om = m\Omega$ and the quantities, $Z^H$, $Z^\infty$
%\footnote{The extra factor of $\sqrt{2\pi}$ comes from normalizing $_{-2}S_{\ell,m}(\theta,\phi)$ to 1.} and $\alpha$ 
%\footnote{The extra factor of 2 comes from calibrating the leading order horizon-absorbed fluxes to agree with those in literature.}
, are given by
\bea
Z^\infty_{\ell,m} &=& \frac{(2\pi)^{3/2}\frak{m}}{2i\omega B_{\ell,m}^\textrm{inc}} \left[ R_{\ell,m}^\textrm{in} \left( A_{nn0} + A_{\bar{m}n0} + A_{\bar{m}\bar{m}0} \right) - R_{\ell,m}^{\textrm{in}\prime} \left( A_{\bar{m}n1} + A_{\bar{m}\bar{m}1} \right) + R_{\ell,m}^{\textrm{in}\prime\prime} A_{\bar{m}\bar{m}2} \right], \nonumber \\
Z^H_{\ell,m} &=&  \frac{(2\pi)^{3/2}\frak{m}B^\textrm{trans}_{\ell,m}}{2i\omega B_{\ell,m}^\textrm{inc}C^\textrm{trans}_{\ell,m}} \left[ R_{\ell,m}^\textrm{out} \left( A_{nn0} + A_{\bar{m}n0} + A_{\bar{m}\bar{m}0} \right) - R_{\ell,m}^{\textrm{out}\prime} \left( A_{\bar{m}n1} + A_{\bar{m}\bar{m}1} \right) + R_{\ell,m}^{\textrm{out}\prime\prime} A_{\bar{m}\bar{m}2} \right], \nonumber \\
\alpha_{\ell,m} &=& \frac{256\,\om^3 k\,(k^2+4\varepsilon^2)(k^2+16\varepsilon^2) (2Mr_+)^5 }{|C_{\ell,m}|^2}.
\eea
Here $R^\textrm{in}_{\ell,m}$ and $R^\textrm{out}_{\ell,m}$ are the homogenous solutions to the ($s=-2$) Teukolsky equation which are ingoing at the future event horizon and outgoing at the future null infinity, respectively, and a prime denotes their derivative with respect to $r$. These solutions and their derivatives are evaluated at $r=r_0$ with an accuracy of more than 600 significant digits, using
\bea
&& R^\textrm{in}_{\ell,m} = e^{i\ep\ka x} (-x)^{2-i\ep_+}(1-x)^{i\ep_-} \sum_{n=-\infty}^\infty a_n F(n+\nu+1-i\tau,-n-\nu-i\tau,3-i\ep-i\tau,x) \nonumber \\
&& R^\textrm{out}_{\ell,m} = 2^\nu e^{-\pi\ep} e^{-i\pi(\nu-1)}e^{i\hat{z}}\hat{z}^{\nu+i\ep_+}(\hat{z}-\ep\ka)^{2-i\ep_+} \nonumber \\
&&\qquad\qquad \times \sum_{n=-\infty}^\infty i^n \frac{ (\nu-1-i\ep)_n }{ (\nu+3+i\ep)_n } a_n (2\hat{z})^n U(n+\nu-1-i\ep,2n+2\nu+2,-2i\hat{z}),
\eea
where 
\bea
&& x = \frac{z_+-z}{\epsilon\kappa},\quad z = \omega r, \quad z_{\pm} = \omega r_{\pm}, \quad \kappa = \sqrt{1-q^2},  \quad q = a/M, \quad \epsilon = 2M\om\nonumber \\
&& \tau = (\epsilon-mq)/\kappa, \quad \epsilon_\pm = (\epsilon\pm\tau)/2, \quad \hat{z} = \omega(r-r_-),\quad r_{\pm} = M \pm \sqrt{M^2-a^2}
\eea
$\nu$, $a_n$ are given in \cite{SasTagLivRev}, and $F$ and $U$ are the hypergeometric and the confluent hypergeometric functions respectively. The remaining quantities are given in Appendix \ref{appA}. 
%%%%%% accuracy %%%%%
The sum-over-$n$ in the above equation have a practical cut-off at some $n_\textrm{max}$; let us denote the homogenous solution for this particular cut-off as $R_{n_\textrm{max}}$ (we have suppressed the indices, (in/out) and $(\ell,m)$, here). By varying $n_\textrm{max}$, we look at the fractional difference between $R_{n_\textrm{max}}$ and $R_{n_\textrm{max}+i}$ (for first few $i$'s from 1 to 10) and choose that $n_\textrm{max}$ where the fractional difference is lower 1 part in $10^{600}$. In Eq(\ref{lmmode}), the sum goes upto $\ell_\textrm{max} = 40$.

\section{Results}
\label{sec3}
In this section, we obtain the analytic coefficients in the pN expansion of the fluxes at the event horizon and infinity in a Kerr spacetime and at the event horizon in a Schwarzschild spacetime, for particle in circular, equatorial orbit. 
%The numerical values are given in tables in Appendix \ref{appB}. 
To numerically extract the spin-dependent pN coefficients, we calculate the fluxes at the horizon and infinity in Kerr and Schwarzschild spacetimes as follows. 

We calculate the flux in Schwarzschild spacetime for 144 different values of $\Omega$ by placing the particle at 
\bea \label{rsvalues}
%r_S/M = 1,2,3,...,9\times 10^{18,19,20,...,33}
R = 1,2,3,...,9\times 10^{18,19,20,...,33}
\eea
where 
%$r_S = M^{1/3}\Omega^{-2/3}$, 
$R = (M\Omega)^{-2/3}$, 
and hence calculate 
%$f_S^{H}(r_{S_i})$ and $f_S^{\infty}(r_{S_i})$ 
$f_S^{H}(R_{i})$ and $f_S^{\infty}(R_{i})$ 
(the index $i$ runs over all the values in Eq.~(\ref{rsvalues})). For a given value of 
%$r_S$ 
$R$ 
in Schwarzschild, we place the particle in Kerr at different radii, $r_K$, for various values of Kerr spin parameters given by
\bea \label{avalues}
a/M = 1,2,3,4,5\times10^{-1,-2,-3,-4,-5}  
\eea
such that
\bea \label{rkrelation}
%r_K = \left( r_S^{3/2} - a M^{1/2} \right)^{2/3},
r_K = \left( (RM)^{3/2} - a M^{1/2} \right)^{2/3},
\eea
and in this way it has the same angular velocity in Kerr as it had in Schwarzschild (when placed at 
%$r_S$
$R$
). This enables us to use a gauge-invariant $R$
%, 
%\bea
%R = \left(M\Omega \right)^{-2/3},
%\eea
 as our expansion parameter. In Kerr spacetime we calculate 3600 different data points for fluxes at infinity, ($f_K^\infty(r_{K_{ij}},a_j)$), and horizon, ($f_K^H(r_{K_{ij}},a_j)$), respectively (the index $j$ runs over all the values in Eq. (\ref{avalues}), the index $i$ runs over all the Eq. (\ref{rsvalues}), and for a given choice of ($i,j$), Eq. (\ref{rkrelation}) is used to evaluate $r_K$). 
We perform a two-dimensional fit by first fitting 
%$[f_K^{H(\infty)}(r_{K_{ij}},a_j) - f_S^{H(\infty)}(r_{S_i})]$
$[f_K^{H(\infty)}(r_{K_{ij}},a_j) - f_S^{H(\infty)}(R_{i})]$
 to a polynomial of the form $c_{i,j,k} a^i \log^j(R)/R^k$ to extract the analytical coefficients, and then subtract them from 
% $[f_K^{H(\infty)}(r_{K_{i,j}},a_j) - f_S^{H(\infty)}(r_{S_i})]$ 
$[f_K^{H(\infty)}(r_{K_{i,j}},a_j) - f_S^{H(\infty)}(R_{i})]$
 and repeat the fitting to extract the numerical coefficients. 
%I have done a similar calculation for the in-falling flux in Schwarzschild (outgoing in Schwarzschild was calculated by Fujita in his famous 22-pN paper) and the in-falling flux in Kerr. The problem for in-falling flux in Kerr is that for a given pN term, there are infinite number of $a$'s, that is, for a given $j$, the j-th pN term is $\sum_{i=1}^\infty \frac{a^i}{R^j}$ unlike for outgoing where the sum over $i$ is finite. Look at  Eq~(4.6) on page 9 of arXiv:gr-qc/9711072v1 to get see of how each pN term looks like. It involves polygamma functions of $a$, $\sqrt{1-a^2}$, etc. 

%We note that $\log^k(R)$ terms starts appearing in the pN expansion of flux from $(1.5+2k)$-pN order onwards at infinity, $3k$-pN onwards for integral pN order and $(3k-0.5)$-pN onwards for half-integral pN order at the horizon. 
We note that $\log^k(R)$ terms first appear at $(1.5+2k)$-pN order at infinity, at $3k$-pN order for integral pN order and at $(3k-0.5)$-pN for half-integral pN order at the horizon.
%The expression given above is accurate upto roughly $3\%$ at $r=6M$. The power series in $a$ converges for all values of $a$ as the magnitude of the absolute values of the coefficients decrease and $a \le 1$ 

\subsection{Flux at infinity in Kerr spacetime}
\label{subsec3}
Here we present the spin-dependent pN coefficients of the flux at infinity in Kerr spacetime. The first quantity on the left-hand side of the equation below is time-averaged flux at infinity in Kerr (represented by the subscript $K$), and the second quantity is the time-averaged flux at infinity in Schwarzschild (represented by the subscript $S$). The pN coefficients of the latter are given in \cite{Fujita}. 
\begin{align} \allowdisplaybreaks
&\left( \left\langle \frac{dE}{dt} \right\rangle_{\textrm{K},\infty} - \left\langle \frac{dE}{dt} \right\rangle_{\textrm{S},\infty} \right) \frac{5R^5}{32} \nonumber \\
&= \left(-\frac{11}{4}a\right) \frac{1}{R^{3/2}} 
% 2 done
+ \left(\frac{33}{16}a^2\right) \frac{1}{R^2} 
% 2.5 done
- \left(\frac{59}{16}a\right)\frac{1}{R^{5/2}}
% 3 done
+ \left(\frac{611}{504}a^2-\frac{65 \pi  }{6}a \right)\frac{1}{R^3}
% 3.5 done
+\left(\frac{162035}{3888}a+\frac{65 \pi  }{8}a^2 -\frac{71 }{24}a^3\right)\frac{1}{R^{7/2}} \nonumber \\
% 4 done
& +\left(-\frac{359 \pi}{14}a +\frac{22667 }{4536}a^2 + \frac{17 }{16}a^4\right)\frac{1}{R^4} 
% 4.5 done
+ \left[\left(-\frac{9828207709}{52390800}+\frac{6841 \gamma}{105}-\frac{43 \pi ^2}{3}+\frac{40939 \log(2)}{315}\right)a  
+ \frac{8447 \pi }{672 }a^2 \right. \nonumber  \\
& \left. -\frac{112025 }{4536 } a^3\right] \frac{1}{R^{9/2}} - \left(\frac{6841}{210}a\right)\frac{\log (R)}{R^{9/2}} 
% 5 done
 + \left[ \frac{23605 \pi }{144}a + \left(\frac{93301799461}{628689600}-\frac{4601 \gamma }{140}+\frac{43 \pi ^2}{4}-\frac{27499
\log (2)}{420} \right)a^2 \right. \nonumber \displaybreak\\
& \left.-\frac{45 \pi }{4} a^3 + \frac{731}{126}a^4\right]\frac{1}{R^{5}} + \left(\frac{4601 }{280}a^2\right)\frac{\log (R)}{R^5} 
% 
% 5.5 done
 + \left[ \left( \frac{-244521688471}{272432160}+\frac{128459 \gamma }{7560}-\frac{671 \pi^2}{12}-\frac{1280791 \log (2)}{10584} \right. \right. \nonumber \\ & \left. \left. +\frac{486243 \log (3)}{3136} \right)a  + \frac{34211 \pi }{1512} a^2 + \frac{-257407}{9072}a^3 + \frac{33 \pi }{8} a^4 + \frac{-1}{8} a^5 \right] \frac{1}{R^{11/2}} +\left(\frac{-128459 }{15120 }a\right) \frac{\log (R)}{R^{11/2}} \nonumber  \\
% 6 done
& \left[ \alpha_{1,6}\,a + \left( \frac{13501670684927}{28605376800}-\frac{22901 \gamma }{196}+\frac{24763 \pi^2}{756}-\frac{537013 \log (2)}{3780}-\frac{142155 \log (3)}{1568} \right) a^2 + \frac{-67426 \pi }{567} a^3 \right. \nonumber \\ & \left. + \frac{24397}{1008} a^4\right] \frac{1}{R^6} + \left(\frac{-40939 \pi }{315}a+ \frac{22901 }{392}a^2 \right) \frac{\log (R)}{R^6} 
% 6.5 done
+ \left(\sum_i \alpha_{i,6.5}a^i\right) \frac{1}{R^{13/2}} + \left(\frac{5045737157}{13471920}a +\frac{27499 \pi   }{420}a^2 \right. \nonumber \\ & \left. -\frac{239171 }{2520 }a^3 \right) \frac{\log(R)}{R^{13/2}}
% 7 done
+ \left[ \alpha_{1,7}\,a + \left( \frac{3708535586671457}{2831932303200}-\frac{921209351 \gamma }{7858620}+\frac{206753\pi ^2}{3402}-\frac{5442958727 \log (2)}{7858620} \right. \right. \nonumber  \\ & \left. \left. + \frac{6830001 \log (3)}{34496} \right)a^2 + \frac{-238825 \pi }{1008} a^3 + \left( \frac{305698147}{1270080}-\frac{1391 \gamma }{84}+\frac{65 \pi ^2}{12}-\frac{4601 \log(2)}{140} \right) a^4 + \frac{-\pi }{4} a^5 + \frac{827}{336} a^6 \right] \frac{1}{R^7}\nonumber \\
& + \left( \frac{-8005397 \pi  }{42336}a   +\frac{921209351 }{15717240}a^2 + \frac{1391}{168}a^4  \right) \frac{\log(R)}{R^7}
% 7.5 done
+ \left( \sum_i \alpha_{i,7.5}\, a^i \right) \frac{1}{R^{15/2}} +  \left[ \left( -\frac{1149838672589069}{471988717200}+\frac{7750438 \gamma }{11025} \right. \right. \nonumber \\ & \left. \left. -\frac{54514 \pi^2}{315}+\frac{9295946 \log (2)}{6615} \right)a + \frac{34227289 \pi }{105840}a^2 + \frac{-105412967}{476280} a^3\right] \frac{\log{r}}{R^{15/2}} + \left( \frac{-3875219}{22050} a\right) \frac{\log^2(R)}{R^{15/2}} \nonumber \\
% 8 done
& + \left( \sum_i \alpha_{i,8}\, a^i \right) \frac{1}{R^8}  + \left[ \frac{14991219653 \pi }{10478160}a + \left( \frac{1644816058929761}{1415966151600}-\frac{652593 \gamma }{2450}+\frac{6099 \pi^2}{70}-\frac{469409 \log (2)}{882} \right)a^2 \right. \nonumber \\
& \left. + \frac{-52971 \pi }{140}a^3 + \frac{2822857}{17640}a^4\right] \frac{\log{r}}{R^8} + \left( \frac{652593}{9800} a^2 \right) \frac{\log^2(R)}{R^8} 
% 8.5 done 
 + \left( \sum_i \alpha_{i,8.5}\, a^i \right) \frac{1}{R^{17/2}} + \left[ \left( -\frac{3095481700672013}{397694196900} \right. \right. \nonumber \\ & \left. \left. + \frac{125764313 \gamma }{1389150}-\frac{22316257\pi ^2}{39690}-\frac{892476173 \log (2)}{463050}+\frac{11571417 \log (3)}{5488} \right)a + \frac{11707148377 \pi }{20956320} a^2 +\frac{-21187719043}{31434480} a^3 \right. \nonumber \\ & \left. + \frac{4601 \pi }{140}a^4 + \frac{-44137}{840}a^5 \right] \frac{\log{r}}{R^{17/2}} + \left(\frac{-125764313}{5556600}a \right) \frac{\log^2(R)}{R^{17/2}}
% 9 done
+ \left( \sum_i \alpha_{i,9}\, a^i \right) \frac{1}{R^{9}} 
+ \bigg[ \beta_{1,9}\,a  \nonumber \\ &  + \left( \frac{11375227513217893139}{1288529197956000} - \frac{1431747488\gamma }{694575}+\frac{4047488 \pi ^2}{6615}-\frac{2160316192\log (2)}{694575}-\frac{5544045 \log (3)}{5488} \right)a^2 \nonumber \\ & +\frac{-467102443 \pi }{381024} a^3 + \frac{820908827}{1164240} a^4\bigg] \frac{\log(R)}{R^9} + \left( \frac{-4647973 \pi }{6615}a + \frac{357936872}{694575}a^2 \right) \frac{\log^2(R)}{R^9}
% 9.5 done
+ \left( \sum_i \alpha_{i,9.5}\, a^i \right) \frac{1}{R^{19/2}}  \nonumber \\ & + \left( \sum_i \beta_{i,9.5}\, a^i \right) \frac{\log(R)}{R^{19/2}} + \left( \frac{127716757532107}{81690354900} a  + \frac{469409 \pi }{1764} a^2 + \frac{-72634489}{66150} a^3 \right) \frac{\log^2(R)}{R^{19/2}} 
% 10 done
+ \left( \sum_i \alpha_{i,10}\, a^i \right) \frac{1}{R^{10}}  \nonumber \\ & +  \left[ \beta_{1,10}\,a + \left( \frac{13001546433395307066857}{542148660039987000}  - \frac{13506324692192 \gamma }{6807529575}+\frac{3740191043 \pi^2}{2143260} -\frac{7490652629792 \log(2)}{972504225} \right.\right. \nonumber \\ & \left. \left. -\frac{52707429 \log (3)}{830060} \right)a^2 + \frac{-10291880279 \pi }{2095632} a^3  + \left( \frac{1132179537999331}{314659144800}-\frac{2942393 \gamma}{22050}+\frac{27499 \pi ^2}{630}-\frac{652593 \log (2)}{2450} \right) a^4 \right.  \nonumber   \\ & \left. + \frac{-12559 \pi }{60}a^5 + \frac{29607}{980} a^6\right] \frac{\log(R)}{R^{10}} + \left( \frac{-4240801783 \pi }{3704400}a +\frac{3376581173048}{6807529575} a^2 + \frac{2942393}{88200} a^4\right) \frac{\log^2(R)}{R^{10}}  \nonumber \\ &
% 10.5 done 
+ \left( \sum_i \alpha_{i,10.5}\, a^i \right) \frac{1}{R^{21/2}} + \left( \sum_i \beta_{i,10.5}\, a^i \right) \frac{\log(R)}{R^{21/2}}  + \left[ \left( -\frac{172422196351100272261}{13287957353921250}+\frac{4135401698\gamma }{1157625}-\frac{30978854 \pi^2}{33075} \right. \right. \nonumber \\
& \left. \left. +\frac{24809960102 \log (2)}{3472875} \right)a + \frac{28509220661 \pi }{11113200} a^2  + \frac{-4224652639}{2000376} a^3\right] \frac{\log^2(R)}{R^{21/2}}  + \left( \frac{-2067700849}{3472875} a\right) \frac{\log^3(R)}{R^{21/2}} \nonumber \\ &
% 11 done
 + \left( \sum_i \alpha_{i,11}\, a^i \right) \frac{1}{R^{11}} + \left( \sum_i \beta_{i,11}\, a^i \right) \frac{\log(R)}{R^{11}} + \left[ \frac{53440237900087 \pi }{10892047320}a + \left( \frac{1287931393290207617089}{159455488247055000}-\frac{836704369 \gamma }{771750} \right. \right. \nonumber \\ & \left. \left. +\frac{7819667 \pi ^2}{22050}-\frac{5019001171\log (2)}{2315250} \right)a^2 + \frac{-16135921 \pi }{3675} a^3 + \frac{15612944489}{5556600} a^4 \right] \frac{\log^2(R)}{R^{11}} + \left( \frac{836704369}{4630500} a^2\right) \frac{\log^3(R)}{R^{11}} \nonumber \displaybreak \\ &
% 11.5 done
+ \left( \sum_i \alpha_{i,11.5}\, a^i \right) \frac{1}{R^{23/2}} + \left( \sum_i \beta_{i,11.5}\, a^i \right) \frac{\log(R)}{R^{23/2}} + \left[ \left( -\frac{36142867921618223066179}{921298376538540000}+\frac{27986513053 \gamma }{13891500}-\frac{1082021803 \pi^2}{297675} \right. \right. \nonumber \\ & \left. \left. -\frac{318932760551 \log(2)}{32413500} +\frac{1065565449 \log (3)}{76832} \right)a + \frac{2513599606631 \pi }{432224100} a^2 + \frac{-89151284892511}{54460236600}a^3 + \frac{652593 \pi }{4900} a^4 \right. \nonumber  \\ & \left. + \frac{-2405467}{3150}a^5 \right] \frac{\log^2(R)}{R^{23/2}}  + \left( -\frac{27986513053}{83349000} a\right) \frac{\log^3(R)}{R^{23/2}} 
% 12 done
 + \left( \sum_i \alpha_{i,12}\, a^i \right) \frac{1}{R^{12}} + \left( \sum_i \beta_{i,12}\, a^i \right) \frac{\log(R)}{R^{12}} + \left[ \gamma_{1,12}\,a \right. \nonumber  \\ & \left. + \left( \frac{3900731059518819150336091}{58041797721928020000} -\frac{416537835481 \gamma }{29172150}+\frac{738463933 \pi^2}{166698}-\frac{1114689921983 \log(2)}{48620250} -\frac{216217755 \log (3)}{38416} \right)a^2 \right. \nonumber  \\ & \left. + \frac{-156203426093 \pi }{14288400} a^3+ \frac{10787551227769}{1210227480} a^4\right] \frac{\log^2(R)}{R^{12}}  + \left( \frac{-24809960102 \pi }{10418625}a + \frac{416537835481}{175032900}a^2 \right) \frac{\log^3(R)}{R^{12}} \nonumber  \\ &
% 12.5 done
+ \left( \sum_i \alpha_{i,12.5}\, a^i \right) \frac{1}{R^{25/2}} + \left( \sum_i \beta_{i,12.5}\, a^i \right) \frac{\log(R)}{R^{25/2}} + \left( \sum_i \gamma_{i,12.5}\, a^i \right) \frac{\log^2(R)}{R^{25/2}} + \left( \frac{95490437041490417}{34660050579000}a \right. \nonumber  \\ & \left. + \frac{5019001171 \pi }{6945750}a^2 +\frac{-346856946221}{41674500}a^3 \right) \frac{\log^3(R)}{R^{25/2}} 
% 13 done
 + \left( \sum_i \alpha_{i,13}\, a^i \right) \frac{1}{R^{13}} + \left( \sum_i \beta_{i,13}\, a^i \right) \frac{\log(R)}{R^{13}} + \left[ \gamma_{1,13}\,a \right. \nonumber \\ & \left. + \left( \frac{18203966121855436121742453919}{110708924974805505348000} -\frac{1000671585632251559 \gamma}{47176179954750}+\frac{366662760623378 \pi^2}{20422588725} -\frac{102994931387593919 \log(2)}{1887047198190}  \right. \right. \nonumber \\ & \left. \left. -\frac{9853498656927 \log (3)}{639146200} \right)a^2 + \frac{-15094769853673 \pi }{550103400} a^3 + \left(  \frac{550687967265477347051}{14173821177516000}-\frac{50226763 \gamma }{92610}+\frac{469409 \pi ^2}{2646}  \right. \right. \nonumber \\
& \left. \left. -\frac{836704369 \log(2)}{771750} \right) a^4 + \frac{-33656957 \pi }{11025} a^5 + \frac{194392307}{111132} a^6\right] \frac{\log^2(R)}{R^{13}} + \left( \frac{-2320552348171 \pi }{388962000}a + \frac{1000671585632251559}{283057079728500} a^2  \right. \nonumber \\
& \left. + \frac{50226763}{555660} a^4 \right)\frac{\log^3(R)}{R^{13}} 
% 13.5 done 
+ \left( \sum_i \alpha_{i,13.5}\, a^i \right) \frac{1}{R^{27/2}} + \left( \sum_i \beta_{i,13.5}\, a^i \right) \frac{\log(R)}{R^{27/2}} + \left( \sum_i \gamma_{i,13.5}\, a^i \right) \frac{\log^2(R)}{R^{27/2}} \nonumber \\ & + \left[ \left( -\frac{4261104058855228415057659}{100456957595644650000}  +\frac{4241868193004 \gamma }{364651875}-\frac{33078313412 \pi^2}{10418625}+\frac{5090136967924 \log (2)}{218791125} \right)a \right. \nonumber \\ & \left. + \frac{6647964457307 \pi }{583443000}a^2 + \frac{-84453761145163}{3938240250}a^3\right]\frac{\log^3(R)}{R^{27/2}} + \left( \frac{-1060467048251}{729303750} a \right)\frac{\log^4(R)}{R^{27/2}} 
% 14 done
 + \left( \sum_i \alpha_{i,14}\, a^i \right) \frac{1}{R^{14}} \nonumber  \\ & + \left( \sum_i \beta_{i,14}\, a^i \right) \frac{\log(R)}{R^{14}} + \left( \sum_i \gamma_{i,14}\, a^i \right) \frac{\log^2(R)}{R^{14}} +\left[ \delta_{1,14}\, a + \left( \frac{2788971974512089547685311217}{43096034808531554850000}-\frac{357978390331 \gamma }{121550625}  \right. \right. \nonumber \\ & \left. \left. +\frac{3345592433 \pi^2}{3472875}-\frac{429547852477 \log (2)}{72930375} \right)a^2  + \frac{-15415403009 \pi }{463050} a^3 + \frac{6787419558527}{218791125} a^4\right]  \frac{\log^3(R)}{R^{14}}  \nonumber \\ & + \left( \frac{357978390331}{972405000}a^2\right) \frac{\log^4(R)}{R^{14}}
% 14.5 done
 + \left( \sum_i \alpha_{i,14.5}\, a^i \right) \frac{1}{R^{29/2}}  + \left( \sum_i \beta_{i,14.5}\, a^i \right) \frac{\log(R)}{R^{29/2}} + \left( \sum_i \gamma_{i,14.5}\, a^i \right) \frac{\log^2(R)}{R^{29/2}}+  \nonumber  \\ & \left( \delta_{1,14.5}\,a+ \frac{710729978088294607 \pi }{18870471981900}a^2 + \frac{2901835701200040089}{56611415945700}a^3 + \frac{836704369 \pi }{2315250}a^4 + \frac{-29284950589}{4630500}a^5 \right) \frac{\log^3(R)}{R^{29/2}}  \nonumber \\ & + \left( -\frac{56894128533671}{26254935000} a \right) \frac{\log^4(R)}{R^{29/2}}
% 15 done
+ \left( \sum_i \alpha_{i,15}\, a^i \right) \frac{1}{R^{15}} + \left( \sum_i \beta_{i,15}\, a^i \right) \frac{\log(R)}{R^{15}} + \left( \sum_i \gamma_{i,15}\, a^i \right) \frac{\log^2(R)}{R^{15}} + \bigg( \delta_{1,15}\,a \nonumber \\ &+\delta_{2,15}\,a^2 + -\frac{3060496172790653 \pi }{31505922000}a^3  + \frac{43045825690358807}{419343821820}a^4 \bigg) \frac{\log^3(R)}{R^{15}} + \left( \frac{-1272534241981 \pi }{218791125}a \right. \nonumber \\ & \left. + \frac{225215586408119}{30630757500}a^2\right) \frac{\log^4(R)}{R^{15}}
% 15.5 done
 + \left( \sum_i \alpha_{i,15.5}\, a^i \right) \frac{1}{R^{31/2}} + \left( \sum_i \beta_{i,15.5}\, a^i \right) \frac{\log(R)}{R^{31/2}} + \left( \sum_i \gamma_{i,15.5}\, a^i \right) \frac{\log^2(R)}{R^{31/2}} \nonumber \\ & + \left( \sum_i \delta_{i,15.5}\, a^i \right) \frac{\log^3(R)}{R^{31/2}} + \left( \frac{-1370058020933843607299}{168135905358729000}a  + \frac{429547852477 \pi }{291721500}a^2 + \frac{-212895380375021}{4375822500}a^3\right) \frac{\log^4(R)}{R^{31/2}} \nonumber \displaybreak \\ &
% 16 done
+ \left( \sum_i \alpha_{i,16}\, a^i \right) \frac{1}{R^{16}} + \left( \sum_i \beta_{i,16}\, a^i \right) \frac{\log(R)}{R^{16}} + \left( \sum_i \gamma_{i,16}\, a^i \right) \frac{\log^2(R)}{R^{16}} +  \bigg( \delta_{1,16}\,a + \delta_{2,16}\,a^2 \nonumber \\ &+ \frac{28510188605919574507 \pi }{377409439638000}a^3 + \delta_{4,16}\,a^4 + \frac{-175689381559 \pi }{6945750}a^5 + \frac{9471843941701}{291721500}a^6 \bigg) \frac{\log^3(R)}{R^{16}} \nonumber \\ & + \left( \frac{-4321875950495813 \pi }{183784545000}a + \frac{3236489054331906396557}{178325960228955000}a^2 + \frac{537033125297}{2917215000} a^4\right)  \frac{\log^4(R)}{R^{16}}
% 16.5 done
+ \left( \sum_i \alpha_{i,16.5}\, a^i \right) \frac{1}{R^{33/2}} \nonumber \\ & + \left( \sum_i \beta_{i,16.5}\, a^i \right) \frac{\log(R)}{R^{33/2}} + \left( \sum_i \gamma_{i,16.5}\, a^i \right) \frac{\log^2(R)}{R^{33/2}} + \left( \sum_i \delta_{i,16.5}\, a^i \right) \frac{\log^3(R)}{R^{33/2}} + \left( \epsilon_{1,16.5} a + \frac{12731208248343149 \pi }{367569090000}a^2 \right. \nonumber \\ & \left. + \frac{-39148960453315463}{206757613125}a^3 \right) \frac{\log^4(R)}{R^{33/2}} + \left( \frac{-529032157531561}{191442234375}a \right)\frac{\log^5(R)}{R^{33/2}}
% 17 done
+ \left( \sum_i \alpha_{i,17}\, a^i \right) \frac{1}{R^{17}} + \left( \sum_i \beta_{i,17}\, a^i \right) \frac{\log(R)}{R^{17}} \nonumber \\ & + \left( \sum_i \gamma_{i,17}\, a^i \right) \frac{\log^2(R)}{R^{17}} + \left( \sum_i \delta_{i,17}\, a^i \right) \frac{\log^3(R)}{R^{17}} + \left( \epsilon_{1,17}\,a + \epsilon_{2,17}\,a^2 + \frac{-3153996932843 \pi }{16206750} a^3 \right. \nonumber \\ & \left. + \frac{59723777854621}{245046060} a^4 \right) \frac{\log^4(R)}{R^{17}} + \left( \frac{51066908514787}{85085437500} a^2 \right) \frac{\log^5(R)}{R^{17}}
% 17.5
+ \left( \sum_i \alpha_{i,17.5}\, a^i \right) \frac{1}{R^{17.5}} + \left( \sum_i \beta_{i,17.5}\, a^i \right) \frac{\log(R)}{R^{17.5}} \nonumber \\ & + \left( \sum_i \gamma_{i,17.5}\, a^i \right) \frac{\log^2(R)}{R^{17.5}} + \left( \sum_i \delta_{i,17.5}\, a^i \right) \frac{\log^3(R)}{R^{17.5}} + \left( \sum_i \epsilon_{i,17.5}\, a^i \right) \frac{\log^4(R)}{R^{17.5}} + \left( \frac{-787775914908495197}{96486886125000} a\right)\frac{\log^5(R)}{R^{17.5}} \nonumber \\
% 18
& + \left( \sum_i \alpha_{i,18}\, a^i \right) \frac{1}{R^{18}} + \left( \sum_i \beta_{i,18}\, a^i \right) \frac{\log(R)}{R^{18}} + \left( \sum_i \gamma_{i,18}\, a^i \right) \frac{\log^2(R)}{R^{18}} + \left( \sum_i \delta_{i,18}\, a^i \right) \frac{\log^3(R)}{R^{18}} + \left( \sum_i \epsilon_{i,18}\, a^i \right) \frac{\log^4(R)}{R^{18}} \nonumber \\ & + \left( \sum_i \zeta_{i,18}\, a^i \right) \frac{\log^5(R)}{R^{18}}
% 18.5
 + \left( \sum_i \alpha_{i,18.5}\, a^i \right) \frac{1}{R^{37/2}} + \left( \sum_i \beta_{i,18.5}\, a^i \right) \frac{\log(R)}{R^{37/2}} + \left( \sum_i \gamma_{i,18.5}\, a^i \right) \frac{\log^2(R)}{R^{37/2}} \nonumber \\ & + \left( \sum_i \delta_{i,18.5}\, a^i \right) \frac{\log^3(R)}{R^{37/2}} + \left( \sum_i \epsilon_{i,18.5}\, a^i \right) \frac{\log^4(R)}{R^{37/2}} + \left( \sum_i \zeta_{i,18.5}\, a^i \right) \frac{\log^5(R)}{R^{37/2}} \nonumber
% 19
+ \left( \sum_i \alpha_{i,19}\, a^i \right) \frac{1}{R^{19}} \nonumber \\ & + \left( \sum_i \beta_{i,19}\, a^i \right) \frac{\log(R)}{R^{19}} + \left( \sum_i \gamma_{i,19}\, a^i \right) \frac{\log^2(R)}{R^{19}} + \left( \sum_i \delta_{i,19}\, a^i \right) \frac{\log^3(R)}{R^{19}} + \left( \sum_i \epsilon_{i,19}\, a^i \right) \frac{\log^4(R)}{R^{19}} \nonumber \\ & + \left( \sum_i \zeta_{i,19}\, a^i \right) \frac{\log^5(R)}{R^{19}} 
% 19.5
 + \left( \sum_i \alpha_{i,19.5}\, a^i \right) \frac{1}{R^{39/2}} + \left( \sum_i \beta_{i,19.5}\, a^i \right) \frac{\log(R)}{R^{39/2}} + \left( \sum_i \gamma_{i,19.5}\, a^i \right) \frac{\log^2(R)}{R^{39/2}} \nonumber \\ & + \left( \sum_i \delta_{i,19.5}\, a^i \right) \frac{\log^3(R)}{R^{39/2}} + \left( \sum_i \epsilon_{i,19.5}\, a^i \right) \frac{\log^4(R)}{R^{39/2}} + \left( \sum_i \zeta_{i,19.5}\, a^i \right) \frac{\log^5(R)}{R^{39/2}} + \left( \sum_i \eta_{i,19.5}\, a^i \right) \frac{\log^6(R)}{R^{39/2}} \nonumber \\ &
 % 120
+ \left( \sum_i \alpha_{i,20}\, a^i \right) \frac{1}{R^{20}} + \left( \sum_i \beta_{i,20}\, a^i \right) \frac{\log(R)}{R^{20}} + \left( \sum_i \gamma_{i,20}\, a^i \right) \frac{\log^2(R)}{R^{20}}  + \left( \sum_i \delta_{i,20}\, a^i \right) \frac{\log^3(R)}{R^{20}} \nonumber \\ & + \left( \sum_i \epsilon_{i,20}\, a^i \right) \frac{\log^4(R)}{R^{20}}  + \left( \sum_i \zeta_{i,20}\, a^i \right) \frac{\log^5(R)}{R^{20}} + \left( \sum_i \eta_{i,20}\, a^i \right) \frac{\log^6(R)}{R^{20}}
\label{i}
\end{align}

\subsection{Flux at the event horizon in Schwarzschild spacetime}
Here we present the flux entering the horizon of the Schwarzschild black hole for a particle in circular orbit.
\begin{align} \allowdisplaybreaks
&\frac{5R^9}{32} \left \langle \frac{dE}{dt} \right \rangle_{S,H}
= \nonumber \\  
%0
&1
%1
+ \frac{4}{R}
%2
+ \left(\frac{172}{7}\right)\frac{1}{R^2}
%3
+ \left(\frac{1272989}{11025}+\frac{16 \pi ^2}{3}-\frac{1712 \log (2)}{105}\right)\frac{1}{R^3} + \left(\frac{1712}{105}\right)\frac{\log(R)}{R^3}
%4
+ \left(\frac{10859497}{22050}-\frac{1024 \gamma }{15}+\frac{52 \pi ^2}{3} \right. \nonumber \displaybreak \\ & \left. -\frac{3980 \log (2)}{21}\right)\frac{1}{R^4} + \left(\frac{9148}{105}\right)\frac{\log(R)}{R^4} 
%5
+ \left(\frac{2547493}{1372}+\frac{2528 \gamma }{35}+\frac{780 \pi ^2}{7}-\frac{425612 \log (2)}{2205}\right)\frac{1}{R^5} + \left(\frac{14732}{49}\right)\frac{\log(R)}{R^5} \nonumber \\ &
% 5.5
+ \left(-\frac{109568 \pi }{1575}\right)\frac{1}{R^{11/2}}
%6
+ \frac{\alpha _6}{R^6} + \left(\frac{1265945848}{694575}+\frac{27392 \pi ^2}{315}-\frac{2930944 \log (2)}{11025}\right)\frac{\log(R)}{R^6}  + \left(\frac{1465472}{11025}\right)\frac{\log^2(R)}{R^6} \nonumber \\ &
%6.5
+ \left(\frac{7239376 \pi }{33075}\right)\frac{1}{R^{13/2}}
%7
+ \frac{\alpha _7}{R^7} + \left(\frac{4017866767}{363825}-\frac{876544 \gamma }{525}+\frac{20176 \pi ^2}{45}-\frac{1059728 \log (2)}{225}\right)\frac{\log(R)}{R^7} + \left(\frac{1736824}{1575}\right)\frac{\log^2(R)}{R^7} \nonumber \\ &
%7.5
+ \left(-\frac{284700044 \pi }{297675}\right)\frac{1}{R^{15/2}}
%8
+ \frac{\alpha _8}{R^8} + \left(\frac{94860587410858}{3476347875}+\frac{6486848 \gamma }{11025}+\frac{3434432 \pi ^2}{2205}-\frac{270364544 \log (2)}{77175}\right)\frac{\log(R)}{R^8} \nonumber \\ & + \left(\frac{56356816}{25725}\right)\frac{\log^2(R)}{R^8}
%8.5
+ \frac{\alpha _{8.5}}{R^{17/2}}  + \left(-\frac{93790208 \pi }{55125}\right)\frac{\log(R)}{R^{17/2}}
%9
+ \frac{\alpha _9}{R^9}+\frac{\beta _9 \log (R)}{R^9} + \left(\frac{5465456790356}{364651875}+\frac{23447552 \pi ^2}{33075} \right. \nonumber \\ & \left. -\frac{2508888064 \log (2)}{1157625}\right) \frac{\log ^2(R)}{R^9} + \left(\frac{2508888064}{3472875}\right)\frac{\log^3(R)}{ R^9} 
%9.5
+ \frac{\alpha _{9.5}}{R^{19/2}} + \left(\frac{6189212128 \pi }{1157625}\right)\frac{\log(R)}{R^{19/2}}
%10
+ \frac{\alpha _{10}}{R^{10}}+\frac{\beta _{10} \log (R)}{R^{10}} \nonumber \\ & + \left(\frac{3756168514560538}{26473726125}-\frac{375160832 \gamma }{18375}+\frac{193425184 \pi ^2}{33075}-\frac{13593351904 \log (2)}{231525}\right) \frac{\log^2(R)}{R^{10}}  + \left(\frac{32514060896}{3472875}\right)\frac{\log^3(R)}{R^{10}} \nonumber \\ &
%10.5
+ \frac{\alpha _{10.5}}{R^{21/2}} + \left(-\frac{243851886344 \pi }{10418625}\right)\frac{\log (R)}{R^{21/2}}
%11
+ \frac{\alpha _{11}}{R^{11}}+\frac{\beta _{11} \log (R)}{R^{11}} + \left(\frac{5274142393747250}{19272872619}-\frac{13707785408 \gamma }{1157625}+\frac{2768388512 \pi ^2}{231525} \right. \nonumber \\ & \left. -\frac{480471905888 \log (2)}{8103375}\right)\frac{\log ^2(R)}{R^{11}} + \left(\frac{336432446912}{24310125}\right) \frac{\log ^3(R)}{R^{11}}
%11.5
+ \frac{\alpha _{11.5}}{R^{23/2}}+\frac{\beta _{11.5} \log (R)}{R^{23/2}}  + \left(-\frac{40142209024 \pi }{1929375}\right)\frac{\log ^2(R)}{R^{23/2}} \nonumber \\ &
%12
+ \frac{\alpha _{12}}{R^{12}}+\frac{\beta _{12} \log (R)}{R^{12}}+\frac{\gamma _{12} \log ^2(R)}{R^{12}} + \left(\frac{7804482398105464}{114865340625}+\frac{40142209024 \pi ^2}{10418625}-\frac{4295216365568 \log (2)}{364651875}\right) \frac{\log ^3(R)}{R^{12}}  \nonumber \\ & + \left(\frac{1073804091392}{364651875}\right)\frac{\log ^4(R)}{R^{12}}
%12.5
+ \frac{\alpha _{12.5}}{R^{25/2}}+\frac{\beta _{12.5} \log (R)}{R^{25/2}} + \left(\frac{2648159561888 \pi }{40516875}\right)\frac{\log ^2(R)}{R^{25/2}}
%13
+ \frac{\alpha _{13}}{R^{13}}+\frac{\beta _{13} \log (R)}{R^{13}} \nonumber \\ & +\frac{\gamma _{13} \log ^2(R)}{R^{13}}  + \left(\frac{1710994261296876104776}{1375971915346875}-\frac{321137672192 \gamma }{1929375}+\frac{519754557824 \pi ^2}{10418625}-\frac{177003777775744 \log(2)}{364651875}\right) \frac{\log ^3(R)}{R^{13}} \nonumber \\ & + \left(\frac{21490311927328}{364651875}\right)\frac{\log ^4(R)}{R^{13}} 
%13.5
+ \frac{\alpha _{13.5}}{R^{27/2}}+\frac{\beta _{13.5} \log (R)}{R^{27/2}} + \left(-\frac{104384512763992 \pi }{364651875}\right)\frac{\log ^2(R)}{R^{27/2}}
%14
+ \frac{\alpha _{14}}{R^{14}}+\frac{\beta _{14} \log (R)}{R^{14}} \nonumber \\ & +\frac{\gamma _{14} \log ^2(R)}{R^{14}}  + \left(\frac{4221703927118507054276}{1626148627228125}-\frac{110527508793728 \gamma }{364651875}+\frac{1868983352192 \pi ^2}{24310125} \right. \nonumber \\ & \left. -\frac{2129235454425088\log (2)}{2552563125}\right) \frac{\log ^3(R)}{R^{14}} + \left(\frac{48431373043184}{510512625}\right)\frac{\log ^4(R)}{R^{14}} 
%14.5
+ \frac{\alpha _{14.5}}{R^{29/2}}+\frac{\beta _{14.5} \log (R)}{R^{29/2}}+\frac{\gamma _{14.5} \log ^2(R)}{R^{29/2}} \nonumber \\ & + \left(-\frac{34361730924544 \pi }{202584375}\right)\frac{\log ^3(R)}{R^{29/2}}
%15
+ \frac{\alpha _{15}}{R^{15}}+\frac{\beta _{15} \log (R)}{R^{15}}+\frac{\gamma _{15} \log ^2(R)}{R^{15}}+\frac{\delta _{15} \log ^3(R)}{R^{15}} + \left(\frac{696250876918789868}{12060860765625} \right. \nonumber \\ & \left. +\frac{17180865462272 \pi ^2}{1093955625}-\frac{1838352604463104 \log (2)}{38288446875}\right) \frac{\log^4(R)}{R^{15}} + \left(\frac{1838352604463104}{191442234375}\right)\frac{\log ^5(R)}{R^{15}}
%15.5 
+ \frac{\alpha _{15.5}}{R^{31/2}}+\frac{\beta _{15.5} \log (R)}{R^{31/2}} \nonumber \\ & +\frac{\gamma _{15.5} \log ^2(R)}{R^{31/2}} + \left(\frac{2266648413992384 \pi }{4254271875}\right)\frac{\log ^3(R)}{R^{31/2}}
%16
+ \frac{\alpha _{16}}{R^{16}}+\frac{\beta _{16} \log (R)}{R^{16}}+\frac{\gamma _{16} \log ^2(R)}{R^{16}}+\frac{\delta _{16} \log ^3(R)}{R^{16}}+\frac{\epsilon _{16}\log ^4(R)}{R^{16}}  \nonumber \\ & + \left(\frac{1148349584799104}{3906984375}\right)\frac{\log ^5(R)}{R^{16}}
%16.5
+ \frac{\alpha _{16.5}}{R^{33/2}}+\frac{\beta _{16.5} \log (R)}{R^{33/2}}+\frac{\gamma _{16.5} \log ^2(R)}{R^{33/2}} + \left(-\frac{89356546683451792 \pi }{38288446875}\right)\frac{\log ^3(R)}{R^{33/2}} \nonumber \\ &
%17
+ \frac{\alpha _{17}}{R^{17}}+\frac{\beta _{17} \log (R)}{R^{17}}  +\frac{\gamma _{17} \log ^2(R)}{R^{17}}+\frac{\delta _{17} \log ^3(R)}{R^{17}}+\frac{\epsilon _{17}\log ^4(R)}{R^{17}} + \left(\frac{178554143515736288}{268019128125}\right)\frac{\log ^5(R)}{R^{17}}
%17.5
+ \frac{\alpha _{17.5}}{R^{35/2}} \nonumber \\ & +\frac{\beta _{17.5} \log (R)}{R^{35/2}}+\frac{\gamma _{17.5} \log ^2(R)}{R^{35/2}}  +\frac{\delta _{17.5} \log ^3(R)}{R^{35/2}} + \left(-\frac{7353410417852416 \pi }{7090453125}\right)\frac{\log ^4(R)}{R^{35/2}}
%18
+ \frac{\alpha _{18}}{R^{18}}+\frac{\beta _{18} \log (R)}{R^{18}}+\frac{\gamma _{18} \log ^2(R)}{R^{18}} \nonumber \\ & +\frac{\delta _{18} \log ^3(R)}{R^{18}}+\frac{\zeta _{18}\log ^5(R)}{R^{18}}+\frac{\epsilon _{18} \log ^4(R)}{R^{18}}   + \left(\frac{1573629829420417024}{60304303828125}\right)\frac{\log ^6(R)}{R^{18}}
%18.5
+ \frac{\alpha _{18.5}}{R^{37/2}}+\frac{\beta _{18.5} \log (R)}{R^{37/2}}+\frac{\gamma _{18.5} \log ^2(R)}{R^{37/2}} \nonumber \\ & +\frac{\delta _{18.5} \log ^3(R)}{R^{37/2}} + \left(\frac{485053335446739872 \pi }{148899515625}\right)\frac{\log ^4(R)}{R^{37/2}} 
%19
+ \frac{\alpha _{19}}{R^{19}} +\frac{\beta _{19} \log (R)}{R^{19}}+\frac{\gamma _{19} \log ^2(R)}{R^{19}}+\frac{\delta _{19} \log ^3(R)}{R^{19}} +\frac{\epsilon _{19} \log ^4(R)}{R^{19}} \nonumber \\ & +\frac{\zeta _{19}\log ^5(R)}{R^{19}} + \left(\frac{73181524665437317376}{60304303828125}\right)\frac{\log ^6(R)}{R^{19}} 
%19.5
+ \frac{\alpha _{19.5}}{R^{39/2}}  +\frac{\beta _{19.5} \log (R)}{R^{39/2}} +\frac{\gamma _{19.5} \log ^2(R)}{R^{39/2}}+\frac{\delta _{19.5} \log^3(R)}{R^{39/2}} \nonumber \\ & +\frac{\epsilon _{19.5} \log ^4(R)}{R^{39/2}}
%20
+ \frac{\alpha _{20}}{R^{20}}+\frac{\beta _{20} \log (R)}{R^{20}}+\frac{\gamma _{20} \log ^2(R)}{R^{20}}   +\frac{\delta _{20} \log ^3(R)}{R^{20}} +\frac{\epsilon _{20} \log ^4(R)}{R^{20}} +\frac{\zeta _{20}\log ^5(R)}{R^{20}}+\frac{\eta _{20} \log ^6(R)}{R^{20}}
%20.5
+ \frac{\alpha _{20.5}}{R^{41/2}} \nonumber \displaybreak \\ & +\frac{\beta _{20.5} \log (R)}{R^{41/2}}+\frac{\gamma _{20.5} \log ^2(R)}{R^{41/2}}+\frac{\delta _{20.5} \log^3(R)}{R^{41/2}}   +\frac{\epsilon _{20.5} \log ^4(R)}{R^{41/2}} +\frac{\zeta _{20.5} \log ^5(R)}{R^{41/2}}
%21
+ \frac{\alpha _{21}}{R^{21}}+\frac{\beta _{21} \log (R)}{R^{21}}+\frac{\gamma _{21} \log ^2(R)}{R^{21}} \nonumber \\ & +\frac{\delta _{21} \log ^3(R)}{R^{21}} +\frac{\epsilon _{21} \log ^4(R)}{R^{21}} +\frac{\zeta _{21}\log ^5(R)}{R^{21}}  +\frac{\eta _{21} \log ^6(R)}{R^{21}}   +\frac{\theta _{21} \log ^7(R)}{R^{21}} 
\label{ii}
\end{align}

\subsection{Flux at the event horizon in Kerr spacetime}
Here we present the pN coefficients of flux entering the event horizon of a Kerr black hole. 
\begin{align} \allowdisplaybreaks \label{iii}
&\left( \left\langle \frac{dE}{dt}\right\rangle_{K,H} - \left\langle \frac{dE}{dt}\right\rangle_{S,H} \right)\frac{5R^{15/2}}{32} \nonumber \\ 
%0.0
&= \left(-\frac{a}{4} -\frac{3 a^3}{4} \right)\frac{1}{R^0} 
%1.0
+ \left( -a -\frac{33 a^3}{16} \right)\frac{1}{R} 
%1.5
+ \left( 2 a B_2+\frac{1}{2}+\frac{13 \kappa  a^2}{2}+\frac{35 a^2}{6}-\frac{a^4}{4} +\frac{\kappa }{2}+3 a^4 \kappa +6 a^3 B_2-1 \right) \frac{1}{R^{3/2}} \nonumber \\ &
%2.0
+ \left( -\frac{43a}{7}-\frac{4651 a^3}{336}-\frac{17 a^5}{56} \right) \frac{1}{R^{2}} 
%2.5
+ \bigg( \frac{433 a^2}{24}-\frac{95 a^4}{24}+2-\frac{3 a^3 B_1}{4}+2 \kappa +\frac{33 a^4 \kappa }{4}+6 a B_2+18 a^3 B_1+\frac{163 \kappa  a^2}{8} +a B_1 \nonumber \\ & -4 \bigg) \frac{1}{R^{5/2}}
%3.0
+ \bigg( -\frac{2586329 a}{44100}-4 B_2-\frac{1640747 a^3}{19600}+19 a^5 \kappa +\frac{428 \gamma a}{105}+\frac{2 \pi ^2 a}{3}+\frac{428}{105} \log (2) a-4 a C_2-12 a^3 C_2  -44 a^2 B_2 \nonumber \\ & +\frac{428 a^3 \gamma }{35}+\frac{428}{35} a^3 \log (2)+2 a^3 \pi ^2+\frac{428}{105} \log (\kappa )a+\frac{428 A_2 a}{105}+\frac{428}{35} a^3 \log (\kappa )+\frac{6 a^7}{\kappa }+\frac{428 a^3A_2}{35}-8 a B_2^2  -24 a^3 B_2^2 \nonumber \\ & -\frac{4 B_2}{\kappa }-\frac{32 a^3}{\kappa }-\frac{31 a}{\kappa}+\frac{57 a^5}{\kappa }+a^4 B_1-\frac{4 a^2 B_1}{3}+\frac{7 a \kappa }{3}+\frac{227 \kappa a^3}{6}+\frac{455 a^5}{16}-\frac{48 a^2 B_2}{\kappa }+\frac{28 a^4 B_2}{\kappa }-\frac{24 a^3C_2}{\kappa }  -\frac{8 a C_2}{\kappa } \nonumber \\ & +\frac{24 a^6 B_2}{\kappa } \bigg) \frac{1}{R^3} + \bigg( -\frac{856 a}{210}-\frac{856 a^3}{70} \bigg) \frac{\log(R)}{R^3}
%3.5
+ \bigg( \frac{19687 a^2}{168}-\frac{145 a^6}{336}-\frac{4729 a^4}{1008}+\frac{899 a B_1}{168}-\frac{41 \kappa  a^6}{28}+\frac{45 a B_3}{56} \nonumber \\ &  -\frac{803 a^3 B_1}{224}+\frac{1665 a^3 B_3}{224}+\frac{719 a^4 \kappa }{12}+\frac{796 a B_2}{21}+\frac{86 \kappa }{7}-\frac{16 a^5 B_2}{7}+\frac{225 a^5 B_3}{28}-\frac{9 a^5 B_1}{28}+\frac{22201 \kappa  a^2}{168}+\frac{2372 a^3 B_2}{21} \nonumber \\ & -\frac{86}{7} \bigg) \frac{1}{R^{7/2}}
%4.0
+ \bigg( -\frac{19366807 a}{88200}-12 B_2-2 B_1-\frac{2062220497 a^3}{6350400}-a C_1+55 a^5 \kappa +\frac{1061 \gamma  a}{35}+\frac{13 \pi ^2 a}{6}+\frac{995}{21} a \log (2) \nonumber \\ & -12 a C_2-36 a^3 C_2+\frac{52 a^4 B_2}{3}-\frac{1136 a^2 B_2}{9}+\frac{12197 a^3 \gamma }{140}+\frac{3873}{28} a^3 \log (2)+\frac{47 a^3 \pi ^2}{8}+\frac{1391}{105} a \log (\kappa )+\frac{428 a A_2}{35} \nonumber \\ & +\frac{5029}{140} a^3 \log (\kappa )+\frac{37 a^7}{6 \kappa }+\frac{1284 a^3 A_2}{35}-24 a B_2^2-72 a^3 B_2^2-\frac{12 B_2}{\kappa }-\frac{341 a^3}{4 \kappa }-\frac{637 a}{6 \kappa }+\frac{741 a^5}{4 \kappa }+\frac{43 a^4 B_1}{6}-\frac{163 a^2 B_1}{18} \nonumber \\ & +\frac{3}{2} a^3 B_1^2+\frac{107 a A_1}{105}-\frac{107 a^3 A_1}{140}-2 a B_1^2+\frac{3 a^3 C_1}{4}-\frac{2 B_1}{\kappa }+\frac{40 \kappa  a}{3}+\frac{815 \kappa  a^3}{6}+\frac{1265 a^5}{18}+\frac{25 a^7}{252}-\frac{144 a^2 B_2}{\kappa }  +\frac{84 a^4 B_2}{\kappa } \nonumber \\ & -\frac{72 a^3 C_2}{\kappa }-\frac{24 a C_2}{\kappa }+\frac{72 a^6 B_2}{\kappa }-\frac{3 a^6 B_1}{\kappa }+\frac{13 a^4 B_1}{2 \kappa }-\frac{3 a^2 B_1}{2 \kappa }-\frac{2 a C_1}{\kappa }+\frac{3 a^3 C_1}{2 \kappa } \bigg) \frac{1}{R^4}+ \biggl( -\frac{4574 a}{210} -\frac{8613 a^3}{140} \biggr)\frac{\log(R)}{R^4} \nonumber \\ &
%4.5
+ \left( \sum_i \alpha_{i,4.5} a^i \right)\frac{1}{R^{9/2}} + \left( \sum_i \beta_{i,4.5} a^i \right)\frac{\log(R)}{R^{9/2}}
%5.0
+ \left( \sum_i \alpha_{i,5} a^i \right)\frac{1}{R^{5}} + \left( -\frac{3683 a}{49}-\frac{471028 a^3}{2205} \right.  \left. -\frac{4867 a^5}{490} \right) \frac{\log(R)}{R^{5}} \nonumber \\ &
%5.5
+ \left( \sum_i \alpha_{i,5.5} a^i \right)\frac{1}{R^{11/2}} + \left( \sum_i \beta_{i,5.5} a^i \right)\frac{\log(R)}{R^{11/2}}
%6.0
+ \left( \sum_i \alpha_{i,6} a^i \right)\frac{1}{R^{6}} + \left( \sum_i \beta_{i,6} a^i \right)\frac{\log(R)}{R^{6}}  + \left( -\frac{366368 a}{11025} \right. \nonumber \\ & \left. -\frac{366368 a^3}{3675} \right) \frac{\log^2(R)}{R^{6}}
%6.5
+ \left( \sum_i \alpha_{i,6.5} a^i \right)\frac{1}{R^{13/2}} + \left( \sum_i \beta_{i,6.5} a^i \right)\frac{\log(R)}{R^{13/2}}
%7.0
+ \left( \sum_i \alpha_{i,7} a^i \right)\frac{1}{R^{7}}  + \left( \sum_i \beta_{i,7} a^i \right)\frac{\log(R)}{R^{7}} \nonumber \\ & + \left( -\frac{434206 a}{1575}-\frac{2007213 a^3}{2450} \right) \frac{\log^2(R)}{R^7}
%7.5
+ \left( \sum_i \alpha_{i,7.5} a^i \right)\frac{1}{R^{15/2}} + \left( \sum_i \beta_{i,7.5} a^i \right)\frac{\log(R)}{R^{15/2}}  + \left( \sum_i \gamma_{i,7.5} a^i \right)\frac{\log^2(R)}{R^{15/2}} \nonumber \\ &
%8.0
+ \left( \sum_i \alpha_{i,8} a^i \right)\frac{1}{R^{8}} + \left( \sum_i \beta_{i,8} a^i \right)\frac{\log(R)}{R^{8}} + \left( -\frac{14089204 a}{25725}-\frac{87225631 a^3}{51450} \right.  \left. -\frac{9849769 a^5}{77175} \right) \frac{\log^2(R)}{R^8} \nonumber \\ &
%8.5
+ \left( \sum_i \alpha_{i,8.5} a^i \right)\frac{1}{R^{17/2}} + \left( \sum_i \beta_{i,8.5} a^i \right)\frac{\log(R)}{R^{17/2}} + \left( \sum_i \gamma_{i,8.5} a^i \right)\frac{\log^2(R)}{R^{17/2}}
%9.0
+ \left( \sum_i \alpha_{i,9} a^i \right)\frac{1}{R^{9}}  + \left( \sum_i \beta_{i,9} a^i \right)\frac{\log(R)}{R^{9}} \nonumber \displaybreak \\ & + \left( \sum_i \gamma_{i,9} a^i \right) \frac{\log^2(R)}{R^9} + \left( -\frac{627222016 a}{3472875}-\frac{627222016 a^3}{1157625} \right) \frac{\log^3(R)}{R^{9}} 
%9.5
+ \left( \sum_i \alpha_{i,9.5} a^i \right)\frac{1}{R^{19/2}}  + \left( \sum_i \beta_{i,9.5} a^i \right)\frac{\log(R)}{R^{19/2}} \nonumber \\ & + \left( \sum_i \gamma_{i,9.5} a^i \right)\frac{\log^2(R)}{R^{19/2}}
%10.0
+ \left( \sum_i \alpha_{i,10} a^i \right)\frac{1}{R^{10}} + \left( \sum_i \beta_{i,10} a^i \right)\frac{\log(R)}{R^{10}} + \left( \sum_i \gamma_{i,10} a^i \right) \frac{\log^2(R)}{R^{10}}   + \left( -\frac{8128515224 a}{3472875} \right. \nonumber \\ & \left. -\frac{2705421598 a^3}{385875} \right) \frac{\log^3(R)}{R^{10}}
%10.5
+ \left( \sum_i \alpha_{i,10.5} a^i \right)\frac{1}{R^{21/2}} + \left( \sum_i \beta_{i,10.5} a^i \right)\frac{\log(R)}{R^{21/2}}  + \left( \sum_i \gamma_{i,10.5} a^i \right)\frac{\log^2(R)}{R^{21/2}} \nonumber \\ & + \left( \sum_i \delta_{i,10.5} a^i \right)\frac{\log^3(R)}{R^{21/2}}
%11.0
+ \left( \sum_i \alpha_{i,11} a^i \right)\frac{1}{R^{11}} + \left( \sum_i \beta_{i,11} a^i \right)\frac{\log(R)}{R^{11}}  + \left( \sum_i \gamma_{i,11} a^i \right) \frac{\log^2(R)}{R^{11}} + \left( -\frac{84108111728 a}{24310125} \right. \nonumber \\ & \left. -\frac{804553210558 a^3}{72930375}-\frac{24989783236 a^5}{24310125} \right) \frac{\log^3{R}}{R^{11}}
%11.5
+ \left( \sum_i \alpha_{i,11.5} a^i \right)\frac{1}{R^{23/2}}  + \left( \sum_i \beta_{i,11.5} a^i \right)\frac{\log(R)}{R^{23/2}} \nonumber \\ & + \left( \sum_i \gamma_{i,11.5} a^i \right)\frac{\log^2(R)}{R^{23/2}} + \left( \sum_i \delta_{i,11.5} a^i \right)\frac{\log^3(R)}{R^{23/2}}
%12.0
+ \left( \sum_i \alpha_{i,12} a^i \right)\frac{1}{R^{12}}  + \left( \sum_i \beta_{i,12} a^i \right)\frac{\log(R)}{R^{12}} \nonumber \\ & + \left( \sum_i \gamma_{i,12} a^i \right) \frac{\log^2(R)}{R^{12}} + \left( \sum_i \delta_{i,12} a^i \right) \frac{\log^3{R}}{R^{12}} + \left(-\frac{268451022848 a}{364651875} \right.  \left. -\frac{268451022848 a^3}{121550625} \right) \frac{\log^4{R}}{R^{12}} \nonumber \\ &
%12.5
+ \left( \sum_i \alpha_{i,12.5} a^i \right)\frac{1}{R^{25/2}} + \left( \sum_i \beta_{i,12.5} a^i \right)\frac{\log(R)}{R^{25/2}} + \left( \sum_i \gamma_{i,12.5} a^i \right)\frac{\log^2(R)}{R^{25/2}}  + \left( \sum_i \delta_{i,12.5} a^i \right)\frac{\log^3(R)}{R^{25/2}} \nonumber \\ &
%13.0
+ \left( \sum_i \alpha_{i,13} a^i \right)\frac{1}{R^{13}} + \left( \sum_i \beta_{i,13} a^i \right)\frac{\log(R)}{R^{13}} + \left( \sum_i \gamma_{i,13} a^i \right) \frac{\log^2(R)}{R^{13}} + \left( \sum_i \delta_{i,13} a^i \right) \frac{\log^3{R}}{R^{13}}  + \left( -\frac{5372577981832 a}{364651875} \right. \nonumber \\ & \left. -\frac{1790422395274 a^3}{40516875} \right) \frac{\log^4{R}}{R^{13}}
%13.5
+ \left( \sum_i \alpha_{i,13.5} a^i \right)\frac{1}{R^{27/2}} + \left( \sum_i \beta_{i,13.5} a^i \right)\frac{\log(R)}{R^{27/2}}  + \left( \sum_i \gamma_{i,13.5} a^i \right)\frac{\log^2(R)}{R^{27/2}} \nonumber \\ & + \left( \sum_i \delta_{i,13.5} a^i \right)\frac{\log^3(R)}{R^{27/2}} + \left( \sum_i \epsilon_{i,13.5} a^i \right)\frac{\log^4(R)}{R^{27/2}}
%14.0
+ \left( \sum_i \alpha_{i,14} a^i \right)\frac{1}{R^{14}} + \left( \sum_i \beta_{i,14} a^i \right)\frac{\log(R)}{R^{14}} \nonumber \\ & + \left( \sum_i \gamma_{i,14} a^i \right) \frac{\log^2(R)}{R^{14}} + \left( \sum_i \delta_{i,14} a^i \right) \frac{\log^3{R}}{R^{14}} + \left( -\frac{12107843260796 a}{510512625}-\frac{575272399991959 a^3}{7657689375} \right. \nonumber \\ & \left. -\frac{1713072165796a^5}{283618125} \right) \frac{\log^4{R}}{R^{14}} 
%14.5
+ \left( \sum_i \alpha_{i,14.5} a^i \right)\frac{1}{R^{29/2}} + \left( \sum_i \beta_{i,14.5} a^i \right)\frac{\log(R)}{R^{29/2}} + \left( \sum_i \gamma_{i,14.5} a^i \right)\frac{\log^2(R)}{R^{29/2}} \nonumber \\ & + \left( \sum_i \delta_{i,14.5} a^i \right)\frac{\log^3(R)}{R^{29/2}} + \left( \sum_i \epsilon_{i,14.5} a^i \right)\frac{\log^4(R)}{R^{29/2}}
%15.0
+ \left( \sum_i \alpha_{i,15} a^i \right)\frac{1}{R^{15}} + \left( \sum_i \beta_{i,15} a^i \right)\frac{\log(R)}{R^{15}} \nonumber \\ & + \left( \sum_i \gamma_{i,15} a^i \right) \frac{\log^2(R)}{R^{15}} + \left( \sum_i \delta_{i,15} a^i \right) \frac{\log^3{R}}{R^{15}}  + \left( \sum_i \epsilon_{i,15} a^i \right) \frac{\log^4{R}}{R^{15}} + \left( -\frac{459588151115776 a}{191442234375} \right. \nonumber \\ & \left. -\frac{459588151115776a^3}{63814078125} \right) \frac{\log^5{R}}{R^{15}}
%15.5
+ \left( \sum_i \alpha_{i,15.5} a^i \right)\frac{1}{R^{31/2}}  + \left( \sum_i \beta_{i,15.5} a^i \right)\frac{\log(R)}{R^{31/2}} + \left( \sum_i \gamma_{i,15.5} a^i \right)\frac{\log^2(R)}{R^{31/2}} \nonumber \\ & + \left( \sum_i \delta_{i,15.5} a^i \right)\frac{\log^3(R)}{R^{31/2}} + \left( \sum_i \epsilon_{i,15.5} a^i \right)\frac{\log^4(R)}{R^{31/2}}
%16.0
+ \left( \sum_i \alpha_{i,16} a^i \right)\frac{1}{R^{16}}  + \left( \sum_i \beta_{i,16} a^i \right)\frac{\log(R)}{R^{16}} \nonumber \\ & + \left( \sum_i \gamma_{i,16} a^i \right) \frac{\log^2(R)}{R^{16}} + \left( \sum_i \delta_{i,16} a^i \right) \frac{\log^3{R}}{R^{16}} + \left( \sum_i \epsilon_{i,16} a^i \right) \frac{\log^4{R}}{R^{16}} + \left( -\frac{287087396199776 a}{3906984375} \right. \nonumber \\ & \left. -\frac{1562969043677416 a^3}{7090453125} \right) \frac{\log^5{R}}{R^{16}}
%16.5
+ \left( \sum_i \alpha_{i,16.5} a^i \right)\frac{1}{R^{33/2}} + \left( \sum_i \beta_{i,16.5} a^i \right)\frac{\log(R)}{R^{33/2}} + \left( \sum_i \gamma_{i,16.5} a^i \right)\frac{\log^2(R)}{R^{33/2}} \nonumber \\ & + \left( \sum_i \delta_{i,16.5} a^i \right)\frac{\log^3(R)}{R^{33/2}} + \left( \sum_i \epsilon_{i,16.5} a^i \right)\frac{\log^4(R)}{R^{33/2}} + \left( \sum_i \zeta_{i,16.5} a^i \right)\frac{\log^5(R)}{R^{33/2}}
%17.0
+ \left( \sum_i \alpha_{i,17} a^i \right)\frac{1}{R^{17}} \nonumber \displaybreak \\ & + \left( \sum_i \beta_{i,17} a^i \right)\frac{\log(R)}{R^{17}}  + \left( \sum_i \gamma_{i,17} a^i \right) \frac{\log^2(R)}{R^{17}} + \left( \sum_i \delta_{i,17} a^i \right) \frac{\log^3{R}}{R^{17}} + \left( \sum_i \epsilon_{i,17} a^i \right) \frac{\log^4{R}}{R^{17}} \nonumber \\ & + \left( -\frac{44638535878934072 a}{268019128125}-\frac{693997312990137382a^3}{1340095640625} \right.  \left. -\frac{37430651315931536 a^5}{1340095640625} \right) \frac{\log^5{R}}{R^{17}}
%17.5
+ \left( \sum_i \alpha_{i,17.5} a^i \right)\frac{1}{R^{35/2}} \nonumber \\ & + \left( \sum_i \beta_{i,17.5} a^i \right)\frac{\log(R)}{R^{35/2}} + \left( \sum_i \gamma_{i,17.5} a^i \right)\frac{\log^2(R)}{R^{35/2}} + \left( \sum_i \delta_{i,17.5} a^i \right)\frac{\log^3(R)}{R^{35/2}} + \left( \sum_i \epsilon_{i,17.5} a^i \right)\frac{\log^4(R)}{R^{35/2}} \nonumber \\ & + \left( \sum_i \zeta_{i,17.5} a^i \right)\frac{\log^5(R)}{R^{35/2}}
%18.0
+ \left( \sum_i \alpha_{i,18} a^i \right)\frac{1}{R^{18}} + \left( \sum_i \beta_{i,18} a^i \right)\frac{\log(R)}{R^{18}}  + \left( \sum_i \gamma_{i,18} a^i \right) \frac{\log^2(R)}{R^{18}} + \left( \sum_i \delta_{i,18} a^i \right) \frac{\log^3{R}}{R^{18}}  \nonumber \\ &  + \left( \sum_i \epsilon_{i,18} a^i \right) \frac{\log^4{R}}{R^{18}} + \left( \sum_i \zeta_{i,18} a^i \right) \frac{\log^5{R}}{R^{18}} + \left( \sum_i \eta_{i,18} a^i \right) \frac{\log^6{R}}{R^{18}} 
%18.5
+ \left( \sum_i \alpha_{i,18.5} a^i \right)\frac{1}{R^{37/2}} + \left( \sum_i \beta_{i,18.5} a^i \right)\frac{\log(R)}{R^{37/2}}  \nonumber \\ & + \left( \sum_i \gamma_{i,18.5} a^i \right)\frac{\log^2(R)}{R^{37/2}} + \left( \sum_i \delta_{i,18.5} a^i \right)\frac{\log^3(R)}{R^{37/2}}  + \left( \sum_i \epsilon_{i,18.5} a^i \right)\frac{\log^4(R)}{R^{37/2}} + \left( \sum_i \zeta_{i,18.5} a^i \right)\frac{\log^5(R)}{R^{37/2}} \nonumber \\ & 
%19.0
+ \left( \sum_i \alpha_{i,19} a^i \right)\frac{1}{R^{19}} + \left( \sum_i \beta_{i,19} a^i \right)\frac{\log(R)}{R^{19}} + \left( \sum_i \gamma_{i,19} a^i \right) \frac{\log^2(R)}{R^{19}} + \left( \sum_i \delta_{i,19} a^i \right) \frac{\log^3{R}}{R^{19}} + \left( \sum_i \epsilon_{i,19} a^i \right) \frac{\log^4{R}}{R^{19}}  \nonumber \\ & + \left( \sum_i \zeta_{i,19} a^i \right) \frac{\log^5{R}}{R^{19}} + \left( \sum_i \eta_{i,19} a^i \right) \frac{\log^6{R}}{R^{19}}
%19.5
+ \left( \sum_i \alpha_{i,19.5} a^i \right)\frac{1}{R^{39/2}}  + \left( \sum_i \beta_{i,19.5} a^i \right)\frac{\log(R)}{R^{39/2}} \nonumber \\ &  + \left( \sum_i \gamma_{i,19.5} a^i \right)\frac{\log^2(R)}{R^{39/2}} + \left( \sum_i \delta_{i,19.5} a^i \right)\frac{\log^3(R)}{R^{39/2}} + \left( \sum_i \epsilon_{i,19.5} a^i \right)\frac{\log^4(R)}{R^{39/2}}  + \left( \sum_i \zeta_{i,19.5} a^i \right)\frac{\log^5(R)}{R^{39/2}}  \nonumber \\ & + \left( \sum_i \eta_{i,19.5} a^i \right)\frac{\log^6(R)}{R^{39/2}}
%20.0
+ \left( \sum_i \alpha_{i,20} a^i \right)\frac{1}{R^{20}} + \left( \sum_i \beta_{i,20} a^i \right)\frac{\log(R)}{R^{20}} + \left( \sum_i \gamma_{i,20} a^i \right) \frac{\log^2(R)}{R^{20}}  + \left( \sum_i \delta_{i,20} a^i \right) \frac{\log^3{R}}{R^{20}}  \nonumber \\ & + \left( \sum_i \epsilon_{i,20} a^i \right) \frac{\log^4{R}}{R^{20}} + \left( \sum_i \zeta_{i,20} a^i \right) \frac{\log^5{R}}{R^{20}} + \left( \sum_i \eta_{i,20} a^i \right) \frac{\log^6{R}}{R^{20}} 
\end{align}
where
\bea
A_n &=& \frac12 \left[ \psi^{(0)}\left( 3 + \frac{niq}{\kappa} \right) + \psi^{(0)}\left( 3 - \frac{niq}{\kappa} \right) \right], \nonumber \\
B_n &=& \frac{1}{2i} \left[ \psi^{(0)}\left( 3 + \frac{niq}{\kappa} \right) - \psi^{(0)}\left( 3 - \frac{niq}{\kappa} \right) \right], \nonumber \\
C_n &=& \frac12 \left[ \psi^{(1)}\left( 3 + \frac{niq}{\kappa} \right) + \psi^{(1)}\left( 3 - \frac{niq}{\kappa} \right) \right], 
\eea
and $\psi^{(n)}(z)$ is the polygamma function.

\section{Discussion}
\begin{figure} 
\centering
\includegraphics[width=0.95\textwidth]{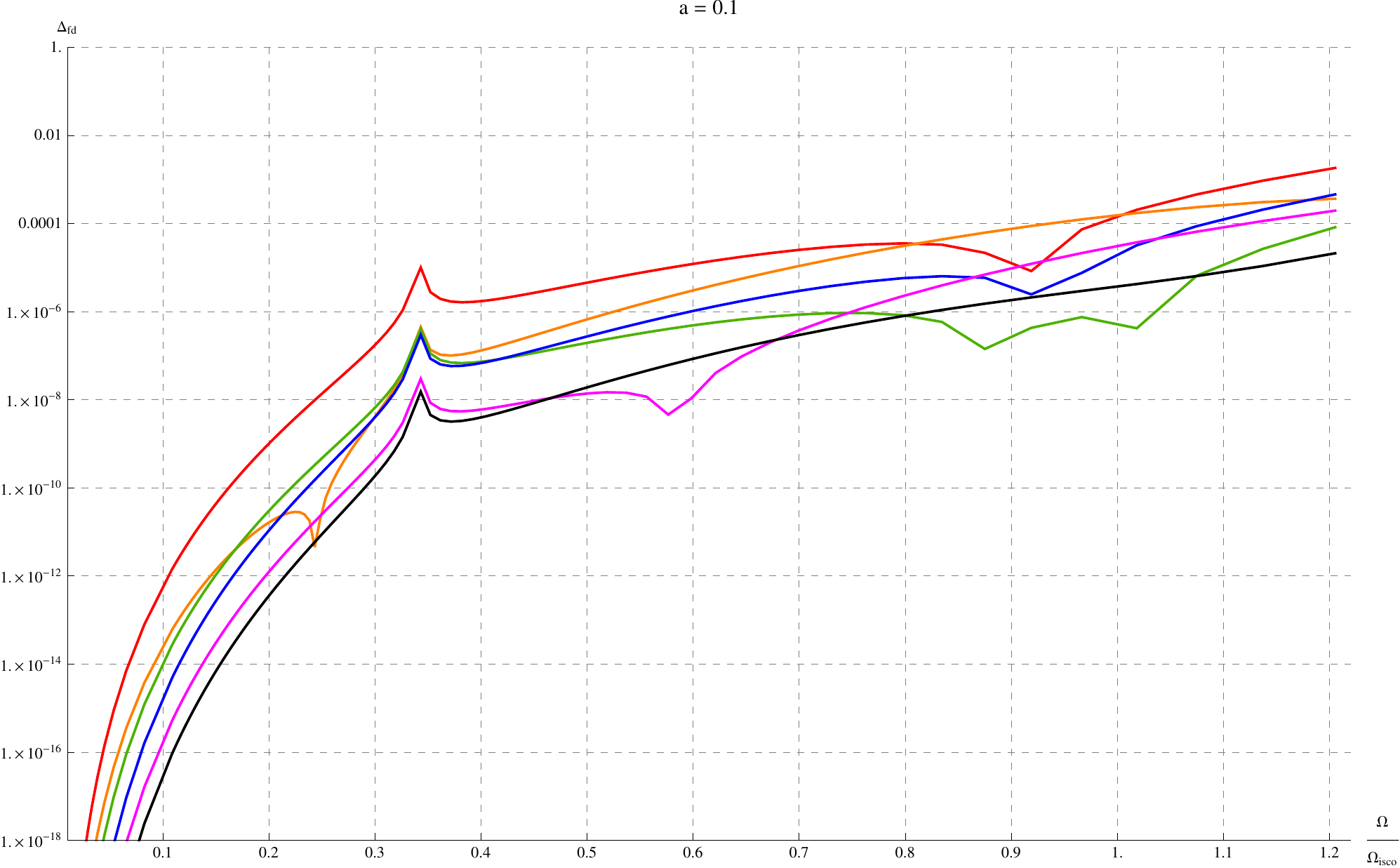}
\par\bigskip\par\bigskip
\includegraphics[width=0.95\textwidth]{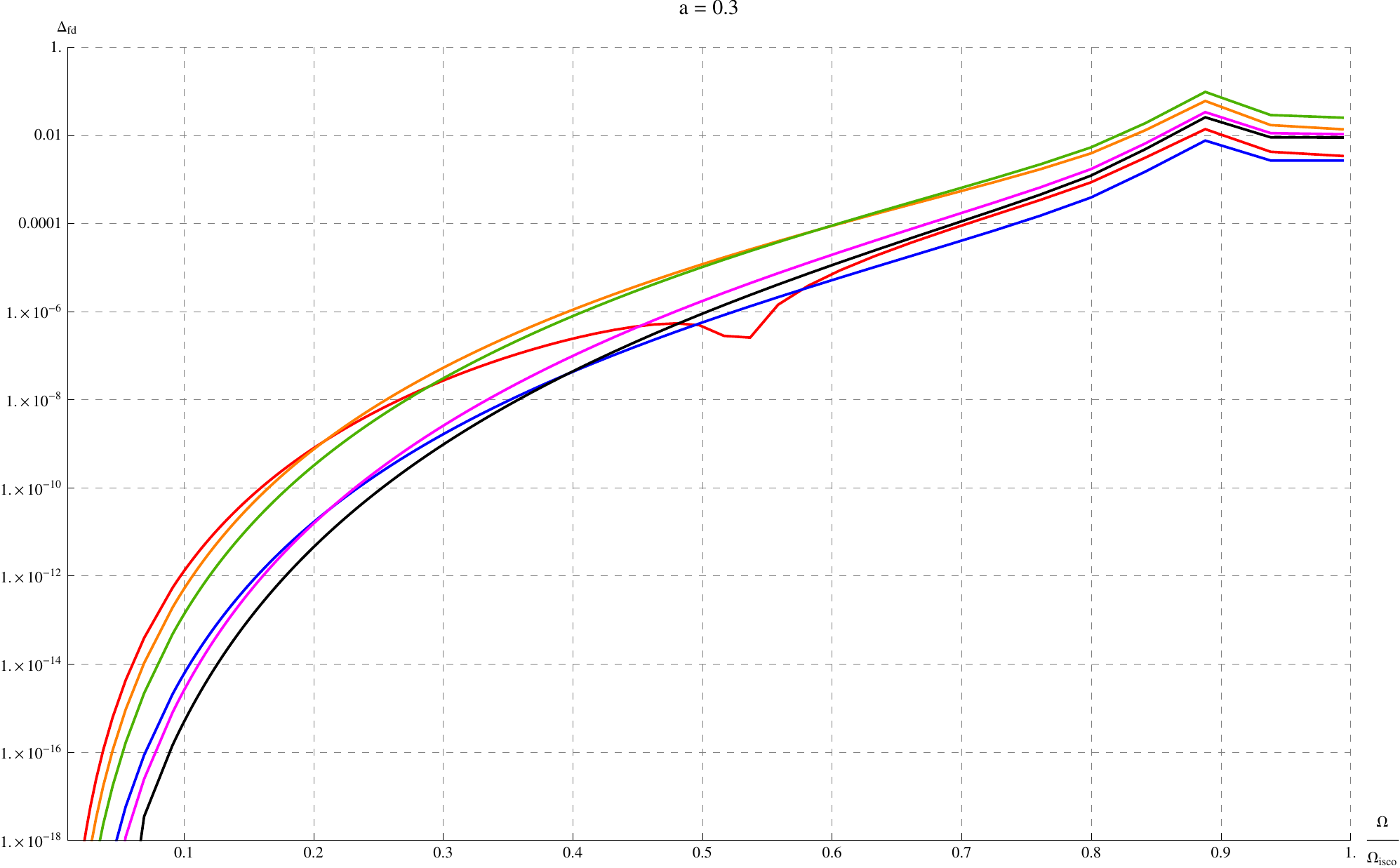}
\caption{Fractional difference ($\Delta_\textrm{fd}$) between the the different pN-order-approximations and the numerical value of the horizon-fluxes at various frequencies compared to the ISCO-frequency for a particular $a$-value. Color-code is as follows - Black: 20pN, Magenta: 19pN, Blue: 18pN, Green: 17pN , Orange: 16pN, Red: 15pN}\label{fig1}
\end{figure}

\begin{figure} 
\centering
\includegraphics[width=0.95\textwidth]{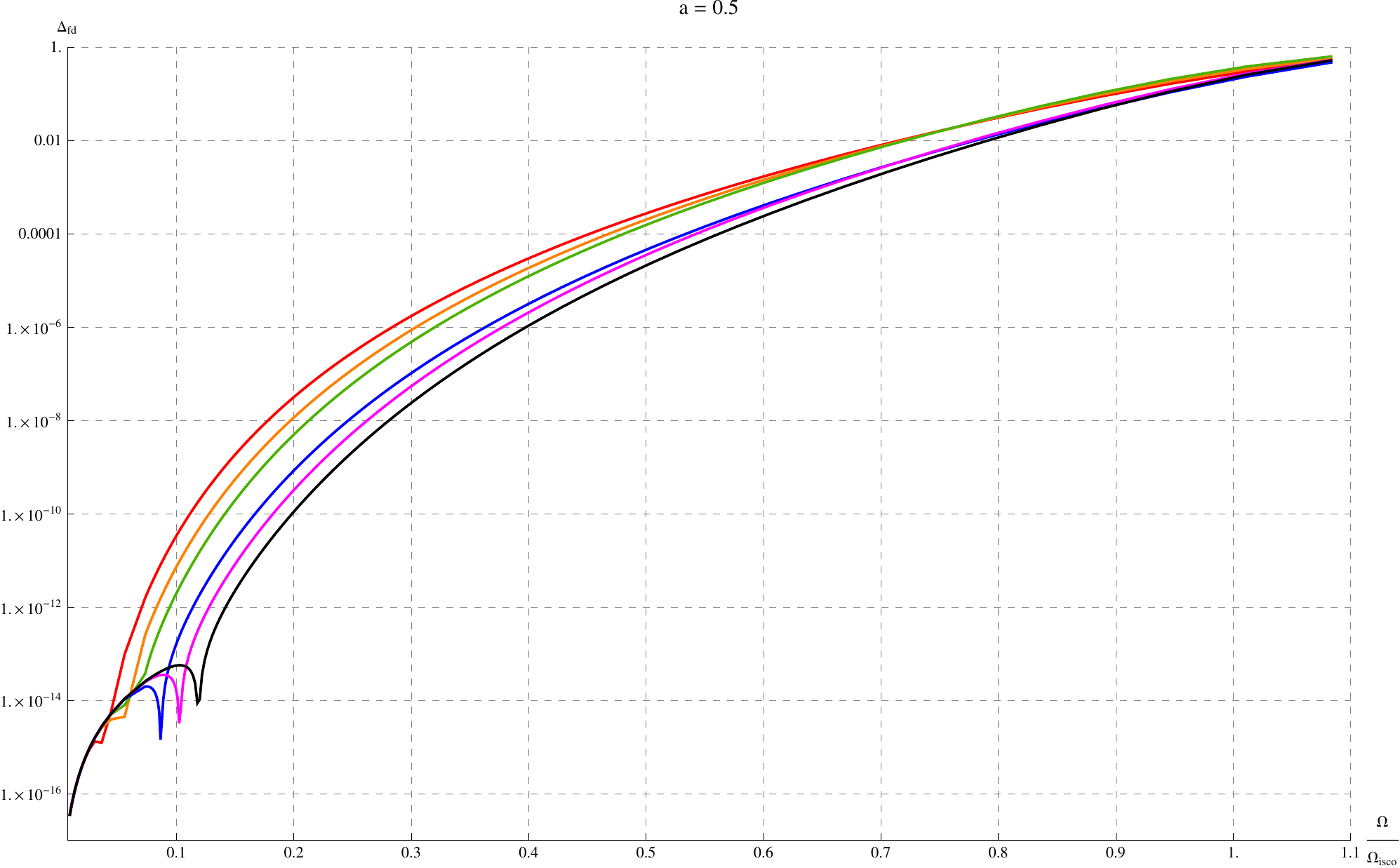}
\par\bigskip\par\bigskip
\includegraphics[width=0.95\textwidth]{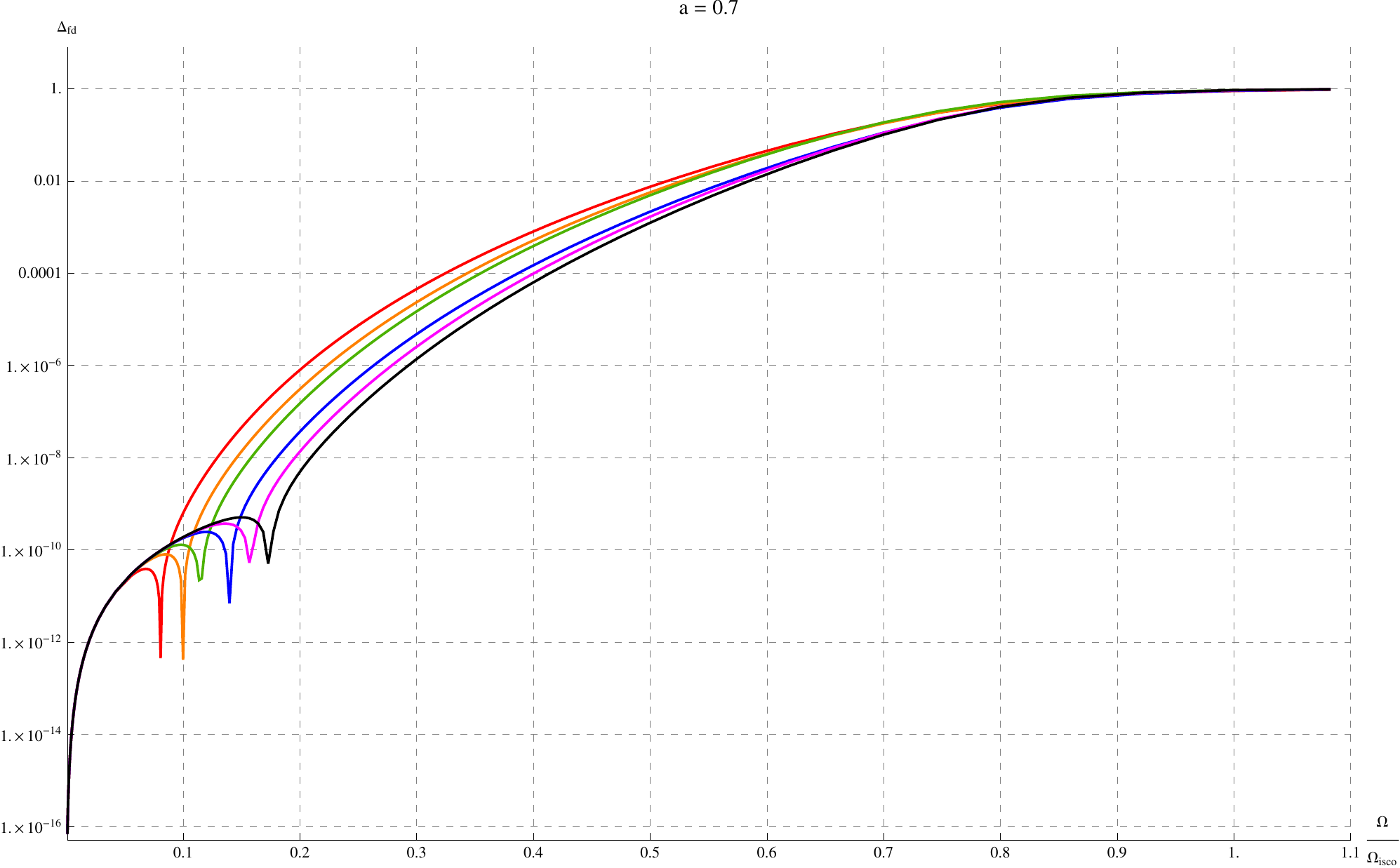}
\caption{Fractional difference ($\Delta_\textrm{fd}$) between the the different pN-order-approximations and the numerical value of the horizon-fluxes at various frequencies compared to the ISCO-frequency for a particular $a$-value. Color-code is as follows - Black: 20pN, Magenta: 19pN, Blue: 18pN, Green: 17pN , Orange: 16pN, Red: 15pN}
\label{fig2}
\end{figure}

\begin{figure}
\centering
\includegraphics[width=0.95\textwidth]{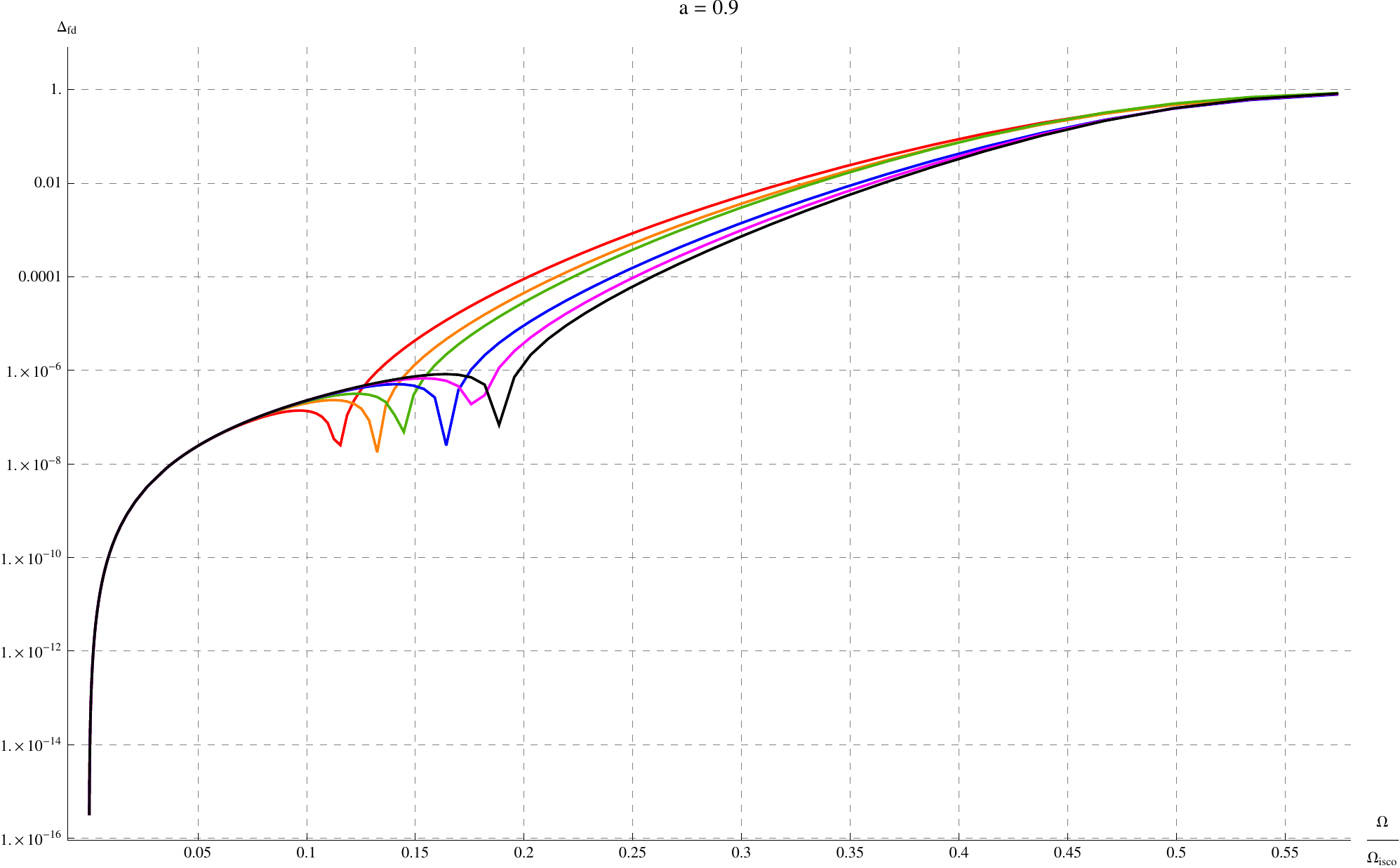}
\par\bigskip\par\bigskip
\includegraphics[width=0.95\textwidth]{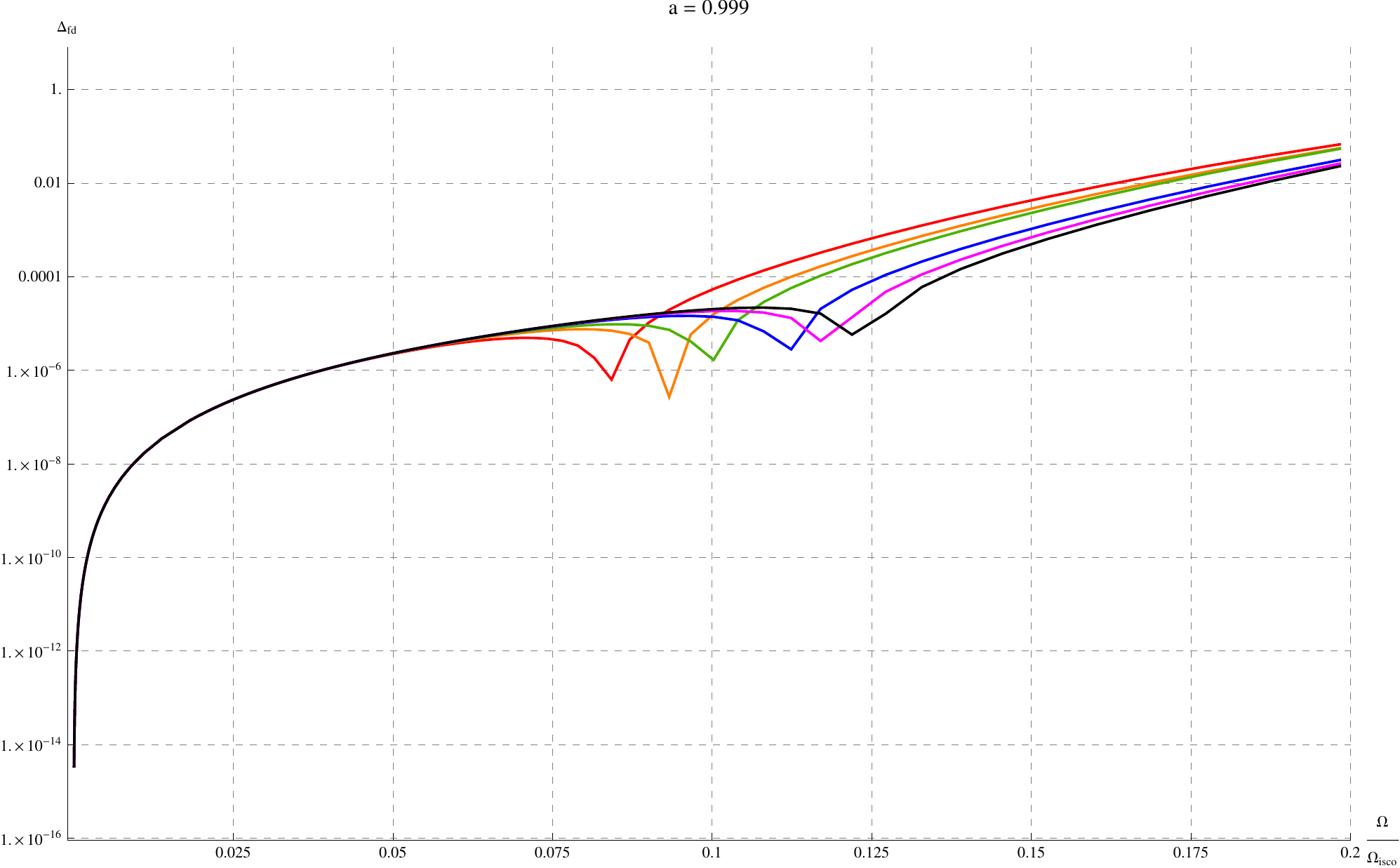}
\caption{Fractional difference ($\Delta_\textrm{fd}$) between the the different pN-order-approximations and the numerical value of the horizon-fluxes at various frequencies compared to the ISCO-frequency for a particular $a$-value. Color-code is as follows - Black: 20pN, Magenta: 19pN, Blue: 18pN, Green: 17pN , Orange: 16pN, Red: 15pN}
\label{fig3}
\end{figure}

\begin{figure}
\centering
\includegraphics[width=0.95\textwidth]{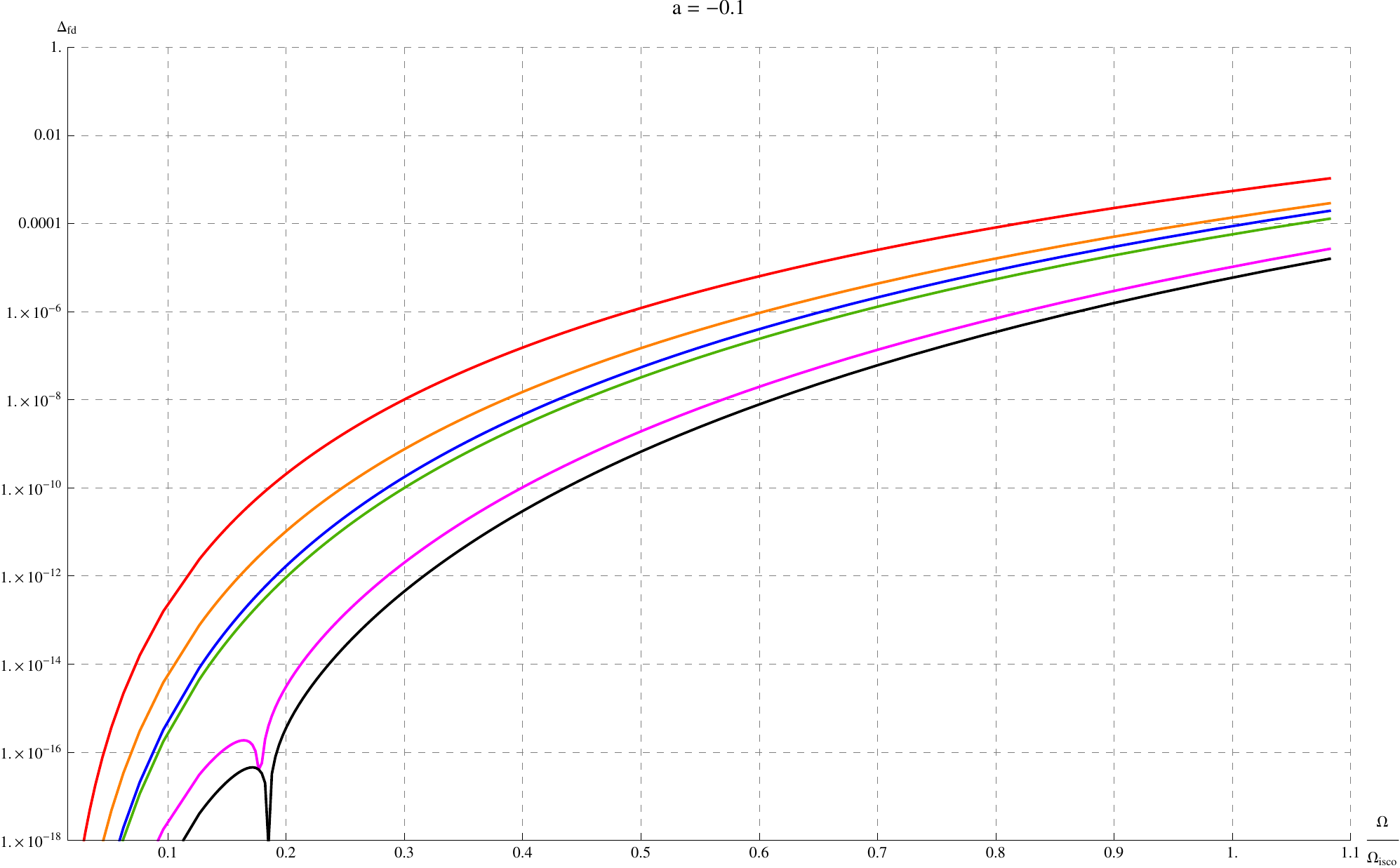}
\par\bigskip\par\bigskip
\includegraphics[width=0.95\textwidth]{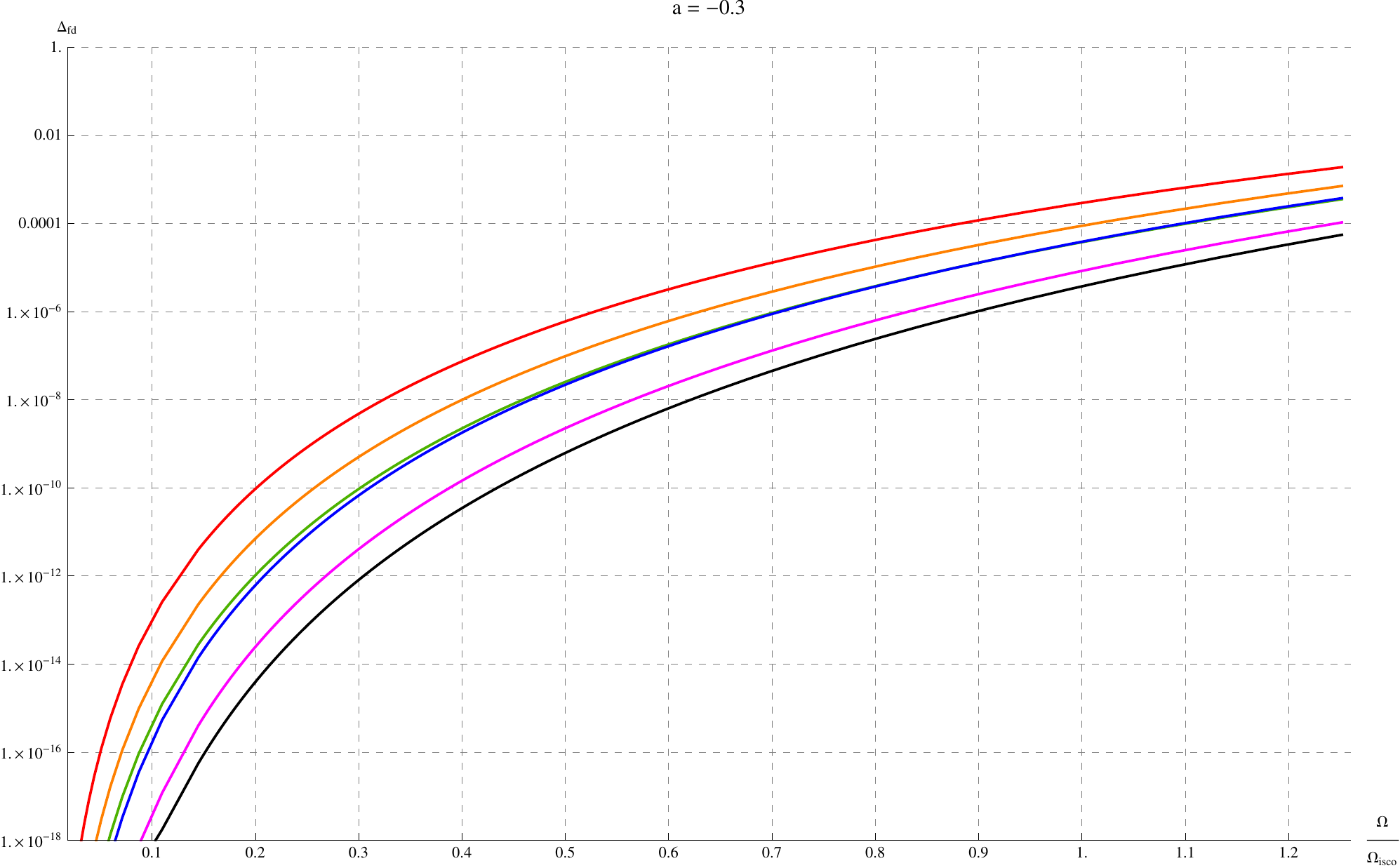}
\caption{Fractional difference ($\Delta_\textrm{fd}$) between the the different pN-order-approximations and the numerical value of the horizon-fluxes at various frequencies compared to the ISCO-frequency for a particular $a$-value. Color-code is as follows - Black: 20pN, Magenta: 19pN, Blue: 18pN, Green: 17pN , Orange: 16pN, Red: 15pN}
\label{fig4}
\end{figure}

\begin{figure}
\centering
\includegraphics[width=0.95\textwidth]{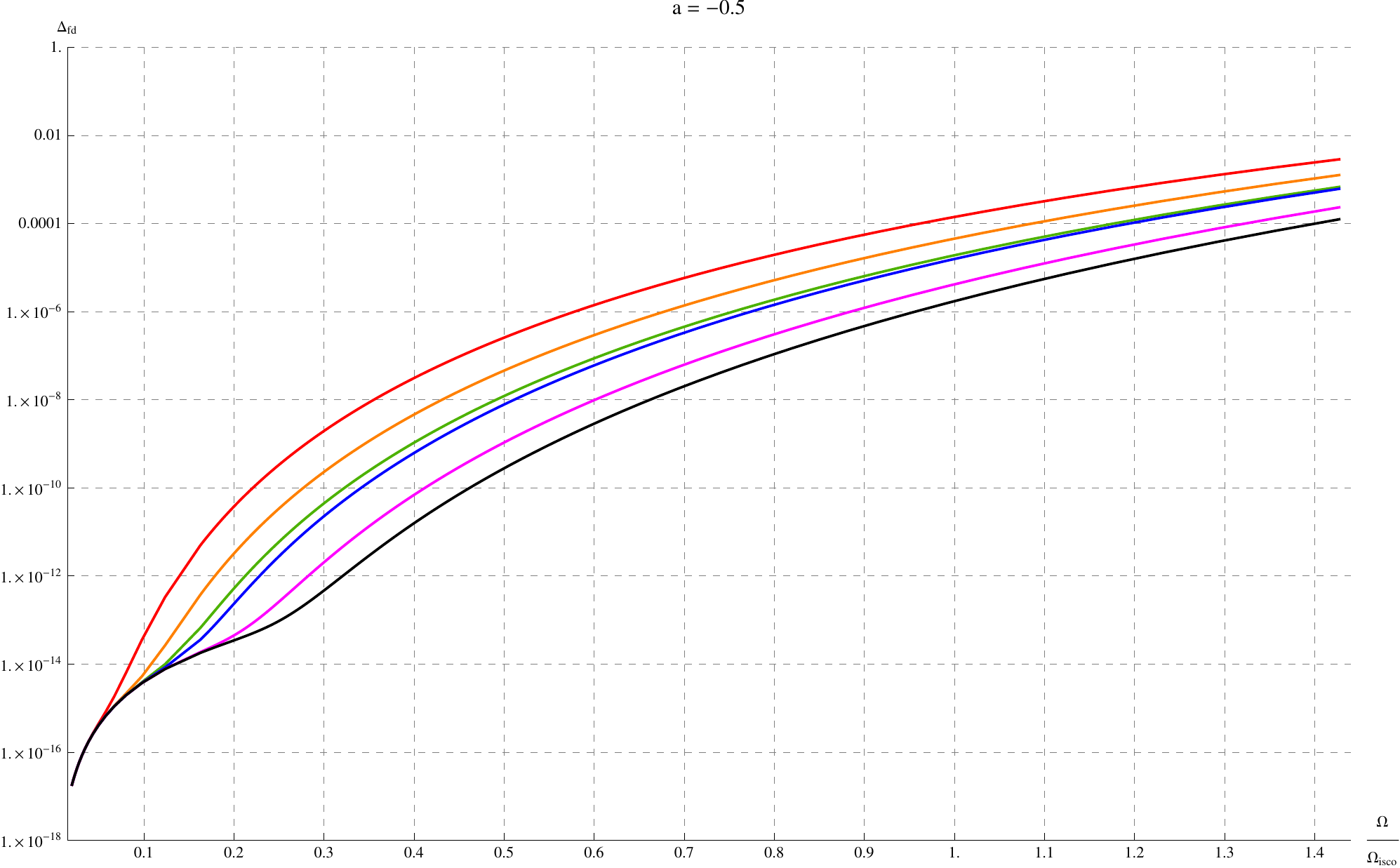}
\par\bigskip\par\bigskip
\includegraphics[width=0.95\textwidth]{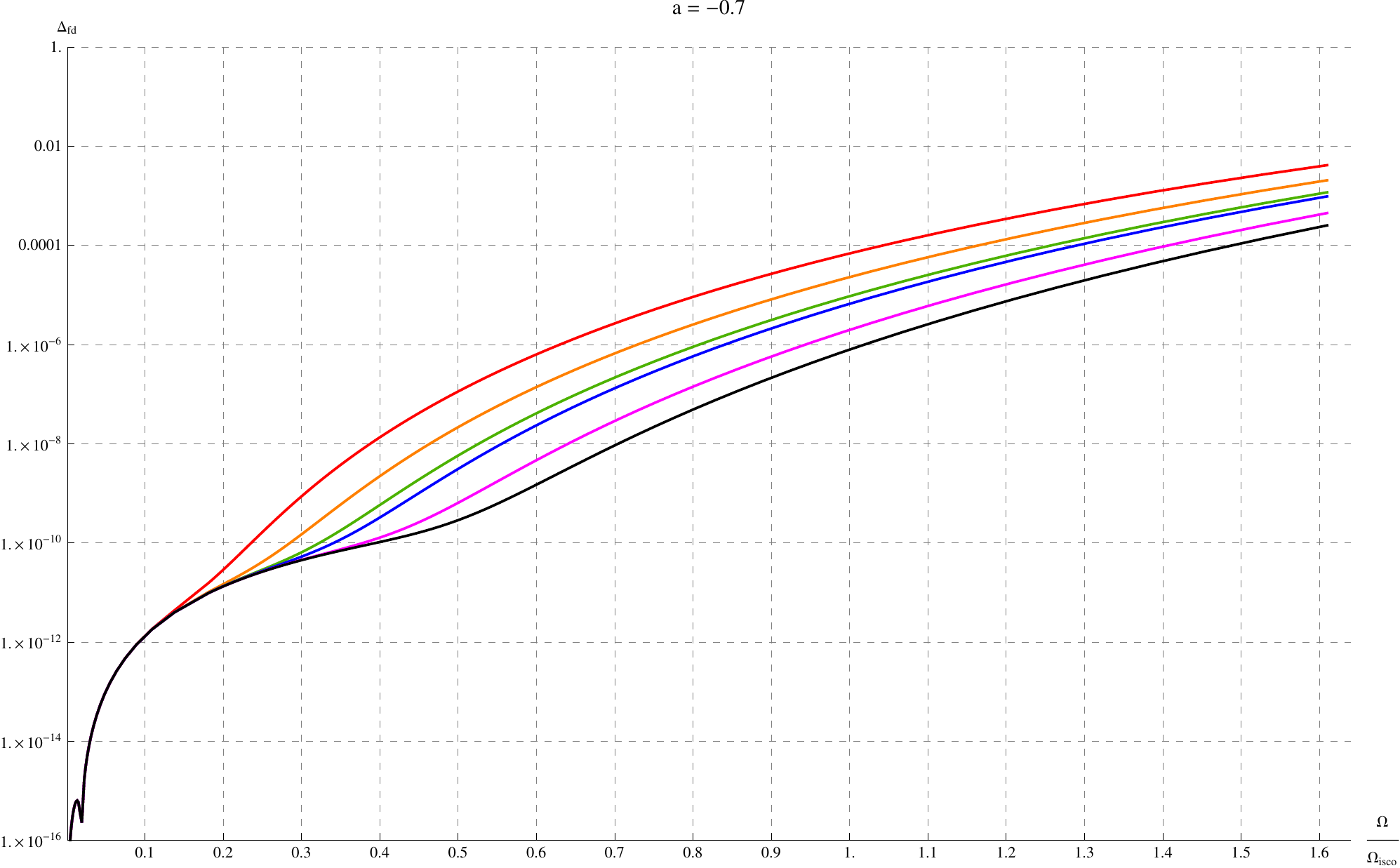}
\caption{Fractional difference ($\Delta_\textrm{fd}$) between the the different pN-order-approximations and the numerical value of the horizon-fluxes at various frequencies compared to the ISCO-frequency for a particular $a$-value. Color-code is as follows - Black: 20pN, Magenta: 19pN, Blue: 18pN, Green: 17pN , Orange: 16pN, Red: 15pN}
\label{fig5}
\end{figure}

\begin{figure}
\centering
\includegraphics[width=0.95\textwidth]{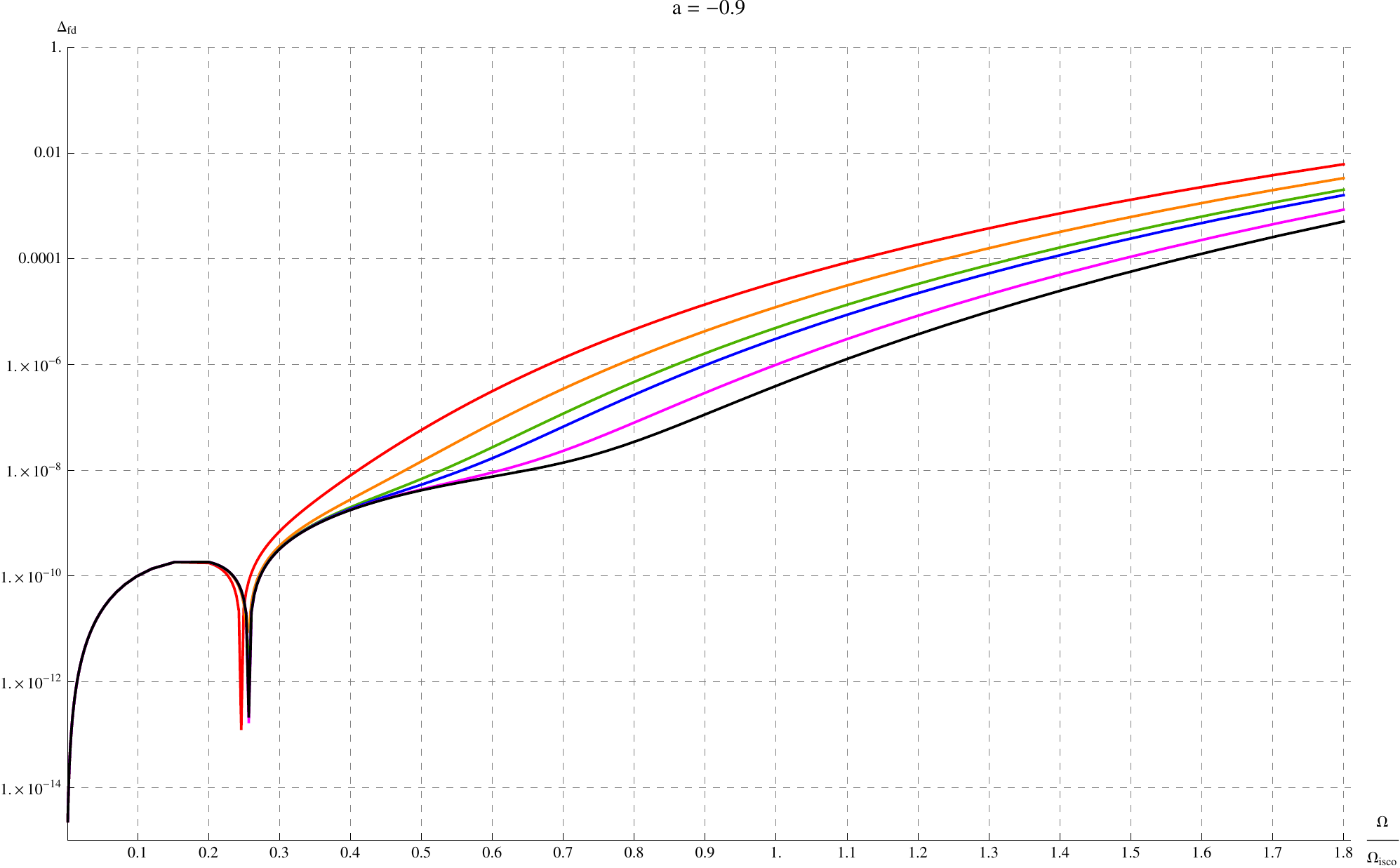}
\par\bigskip\par\bigskip
\includegraphics[width=0.95\textwidth]{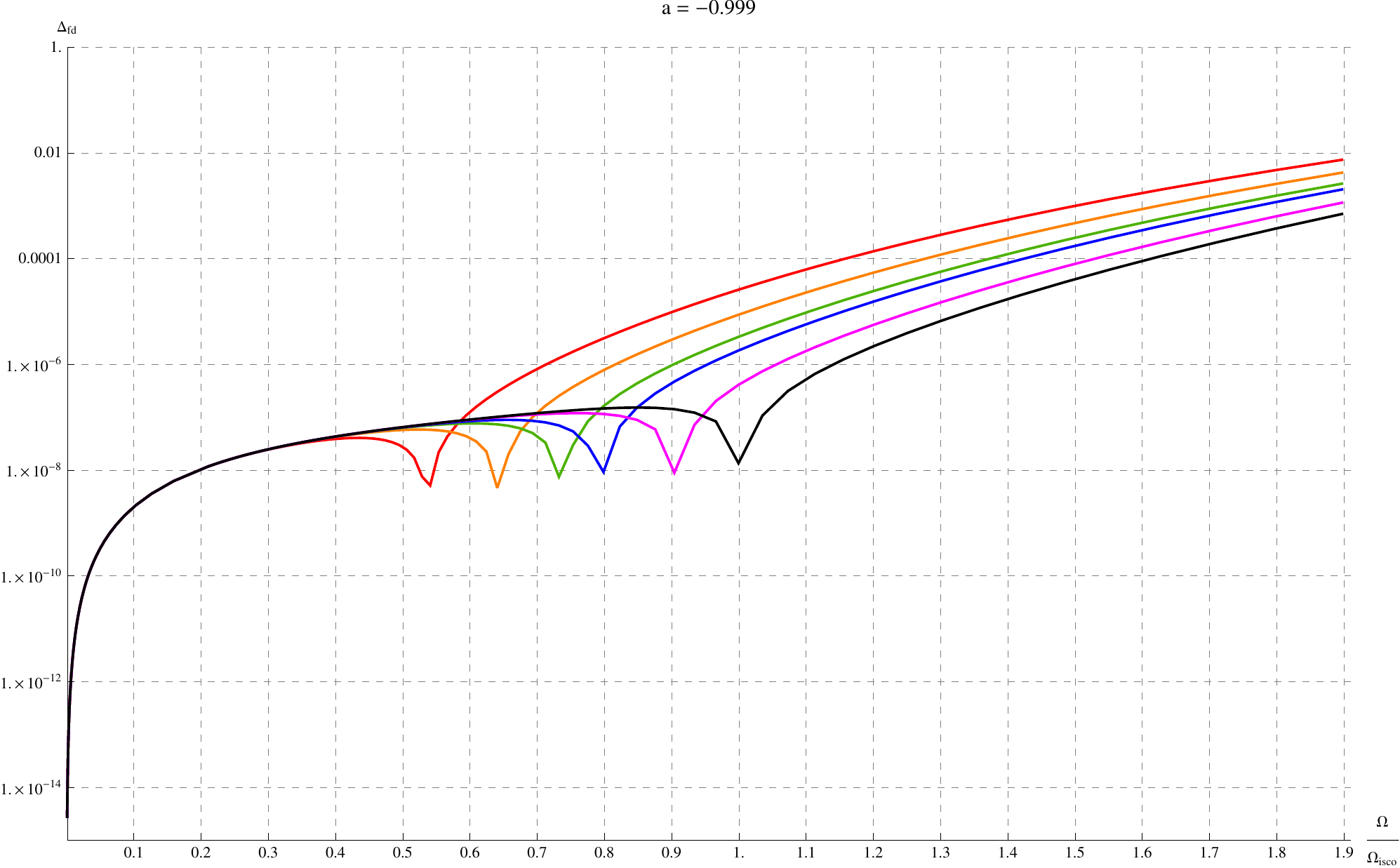}
\caption{Fractional difference ($\Delta_\textrm{fd}$) between the the different pN-order-approximations and the numerical value of the horizon-fluxes at various frequencies compared to the ISCO-frequency for a particular $a$-value. Color-code is as follows - Black: 20pN, Magenta: 19pN, Blue: 18pN, Green: 17pN , Orange: 16pN, Red: 15pN}
\label{fig6}
\end{figure}

\begin{figure}
\centering
\includegraphics[width=0.95\textwidth]{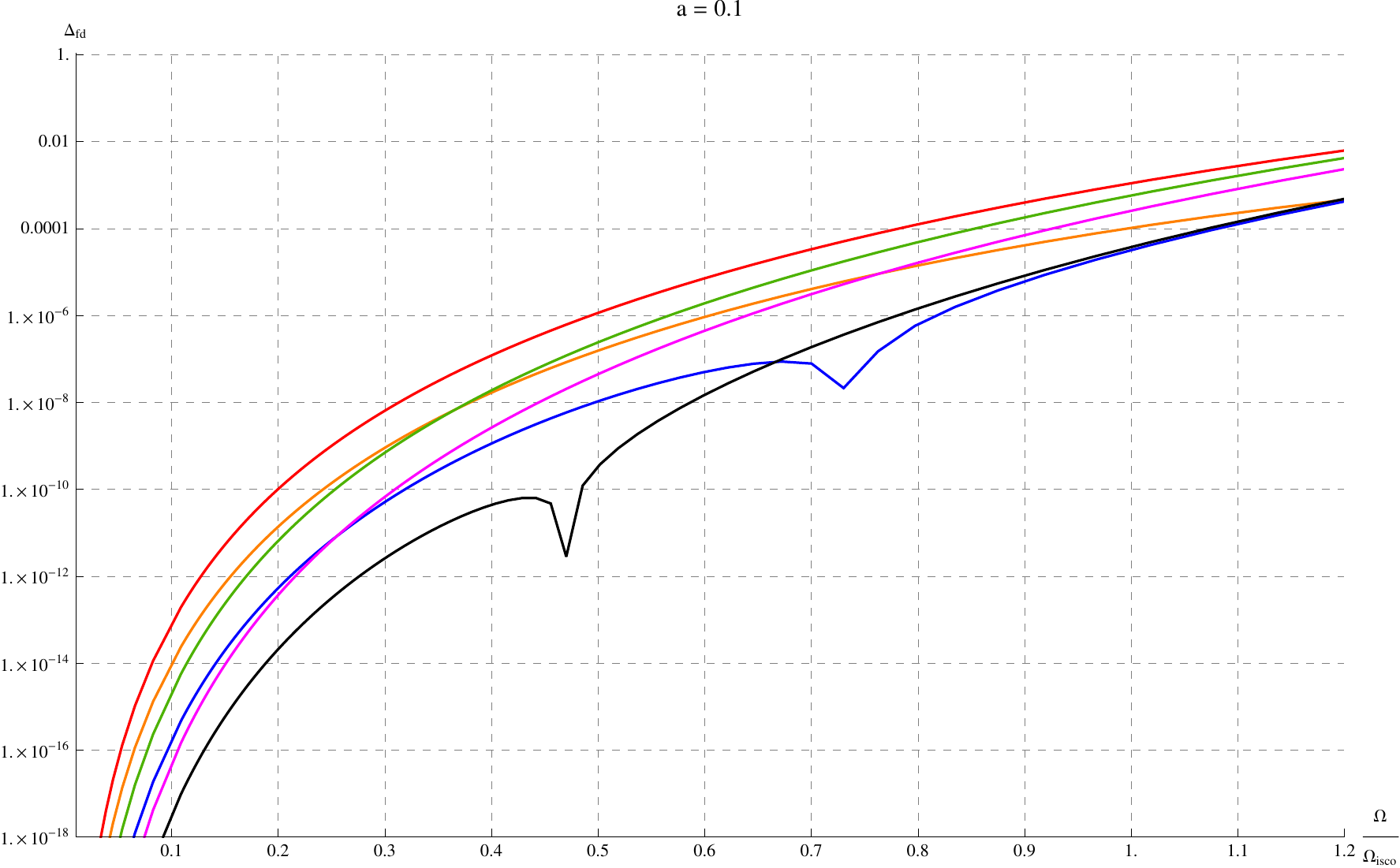}
\par\bigskip\par\bigskip
\includegraphics[width=0.95\textwidth]{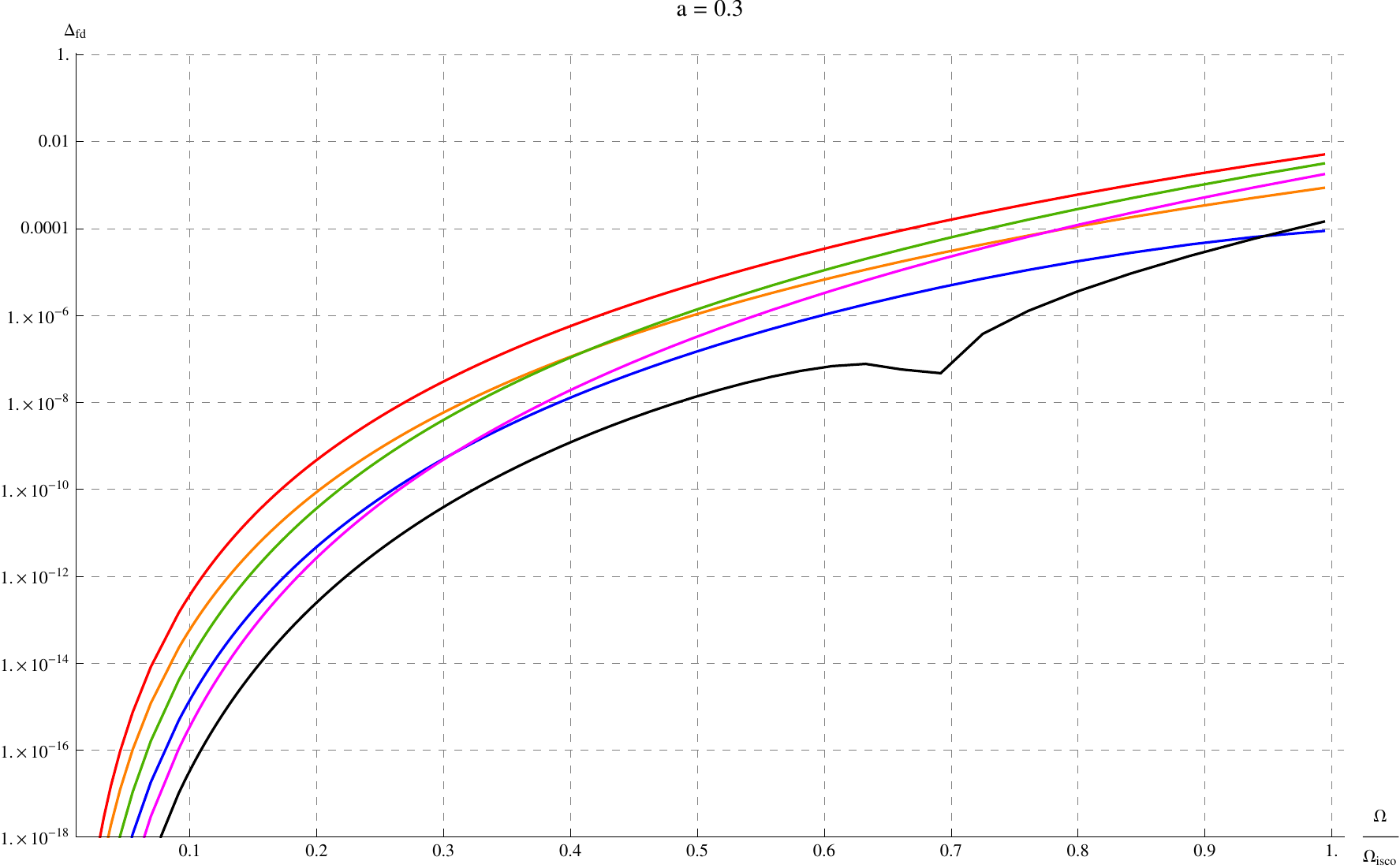}
\caption{Fractional difference ($\Delta_\textrm{fd}$) between the the different pN-order-approximations and the numerical value of the infinity-fluxes at various frequencies compared to the ISCO-frequency for a particular $a$-value. Color-code is as follows - Black: 20pN, Magenta: 19pN, Blue: 18pN, Green: 17pN , Orange: 16pN, Red: 15pN}
\label{fig7}
\end{figure}

\begin{figure}
\centering
\includegraphics[width=0.95\textwidth]{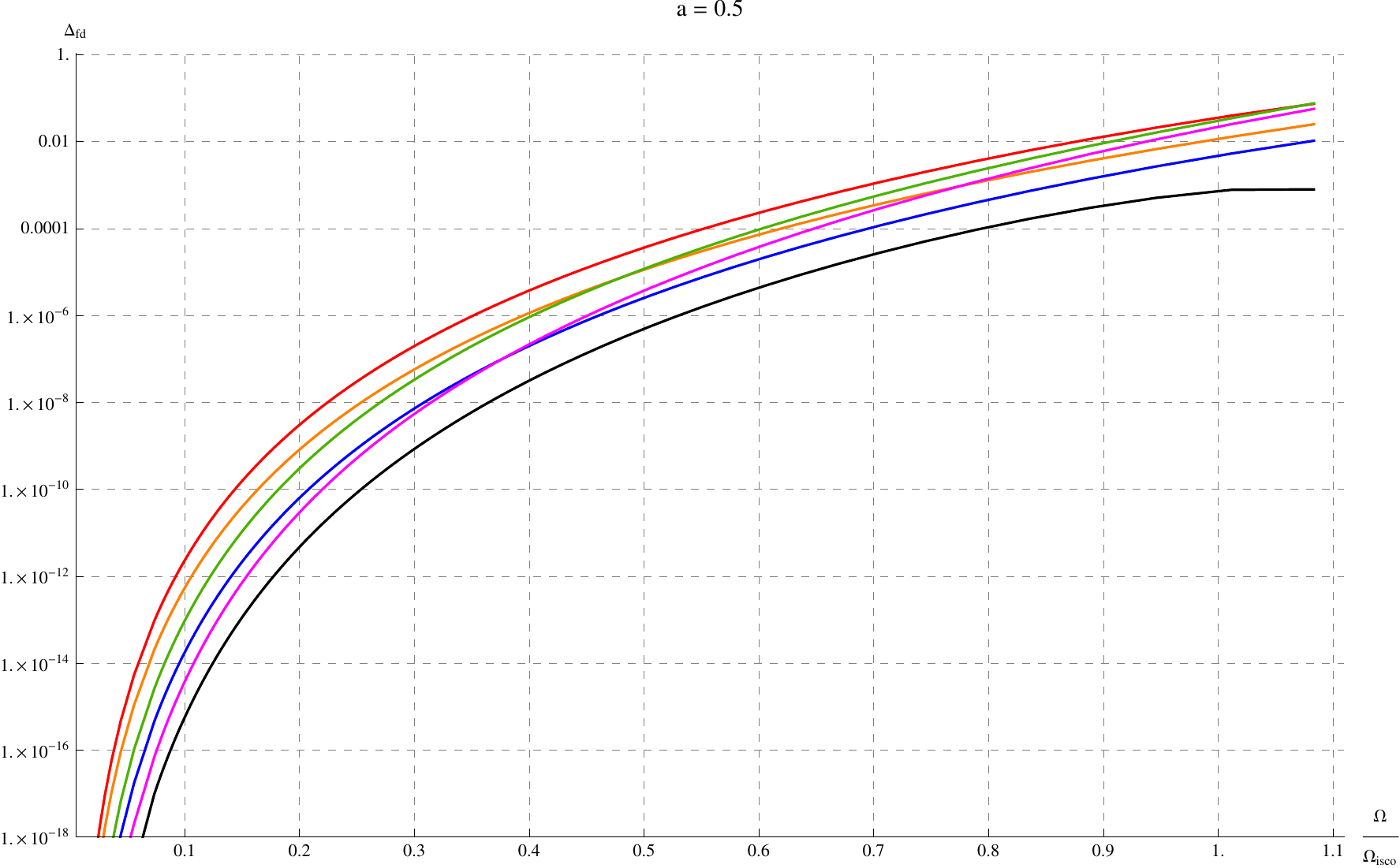}
\par\bigskip\par\bigskip
\includegraphics[width=0.95\textwidth]{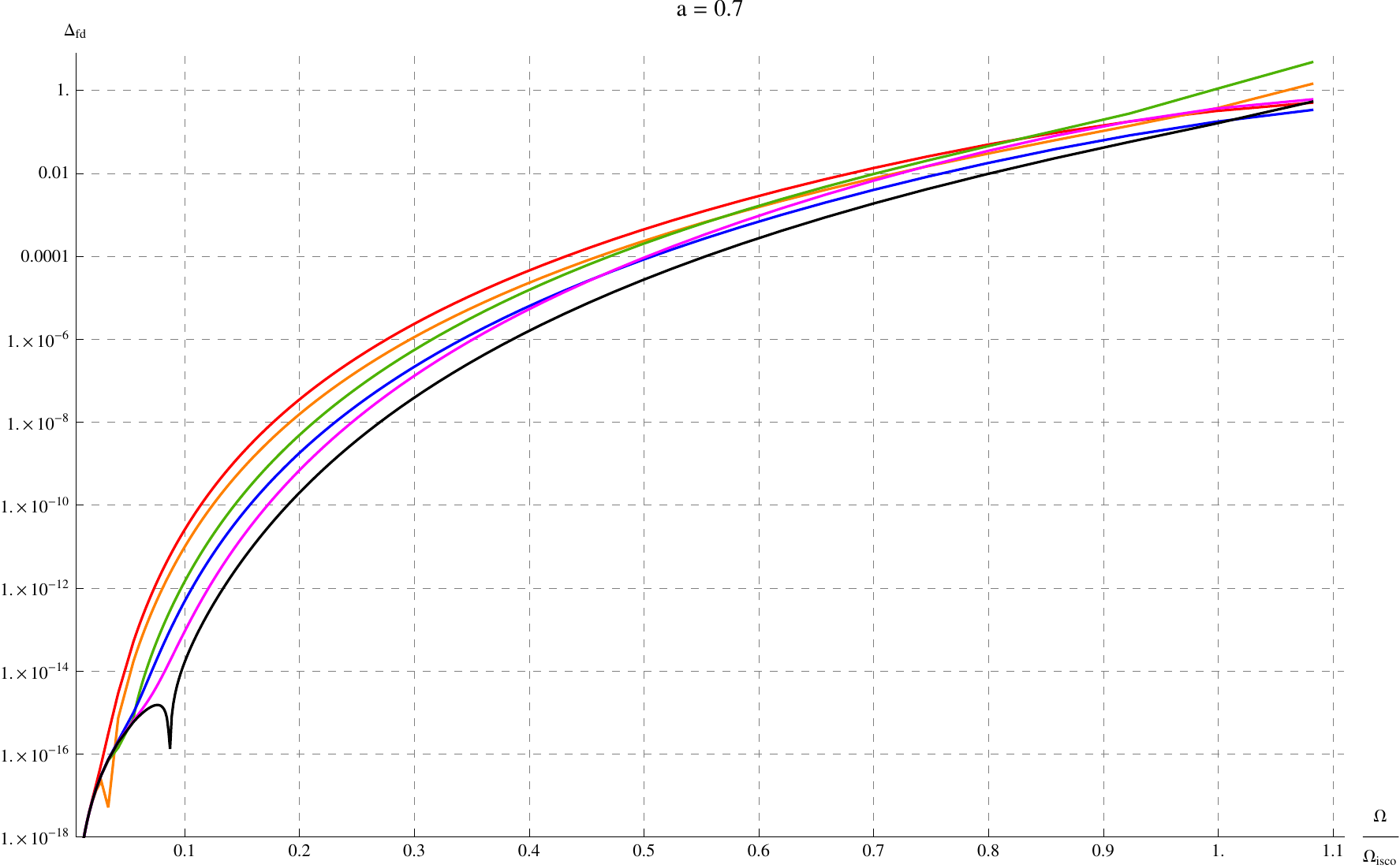}
\caption{Fractional difference ($\Delta_\textrm{fd}$) between the the different pN-order-approximations and the numerical value of the infinity-fluxes at various frequencies compared to the ISCO-frequency for a particular $a$-value. Color-code is as follows - Black: 20pN, Magenta: 19pN, Blue: 18pN, Green: 17pN , Orange: 16pN, Red: 15pN}
\label{fig8}
\end{figure}

\begin{figure}
\centering
\includegraphics[width=0.95\textwidth]{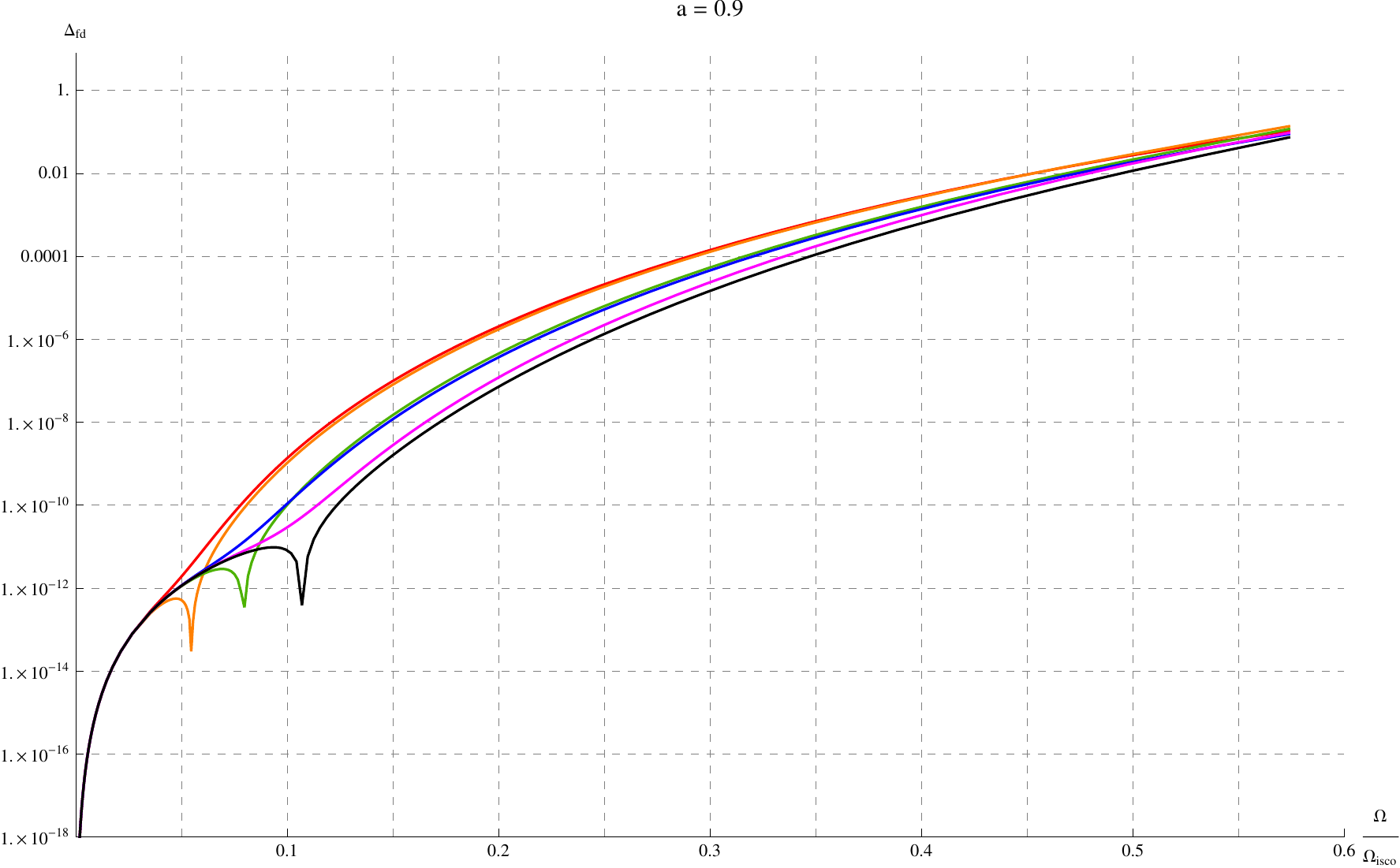}
\par\bigskip\par\bigskip
\includegraphics[width=0.95\textwidth]{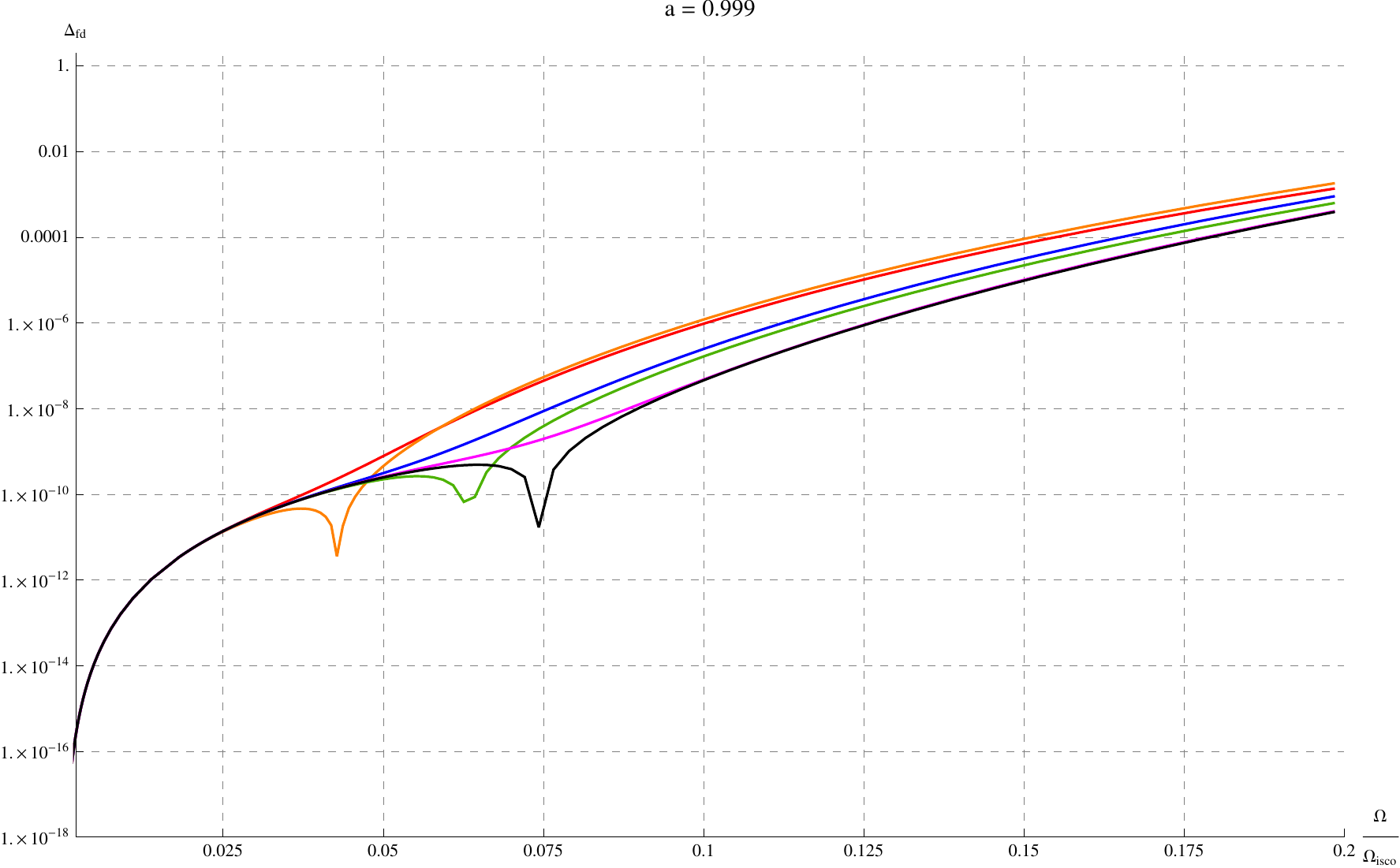}
\caption{Fractional difference ($\Delta_\textrm{fd}$) between the the different pN-order-approximations and the numerical value of the infinity-fluxes at various frequencies compared to the ISCO-frequency for a particular $a$-value. Color-code is as follows - Black: 20pN, Magenta: 19pN, Blue: 18pN, Green: 17pN , Orange: 16pN, Red: 15pN}
\label{fig9}
\end{figure}

\begin{figure}
\centering
\includegraphics[width=0.95\textwidth]{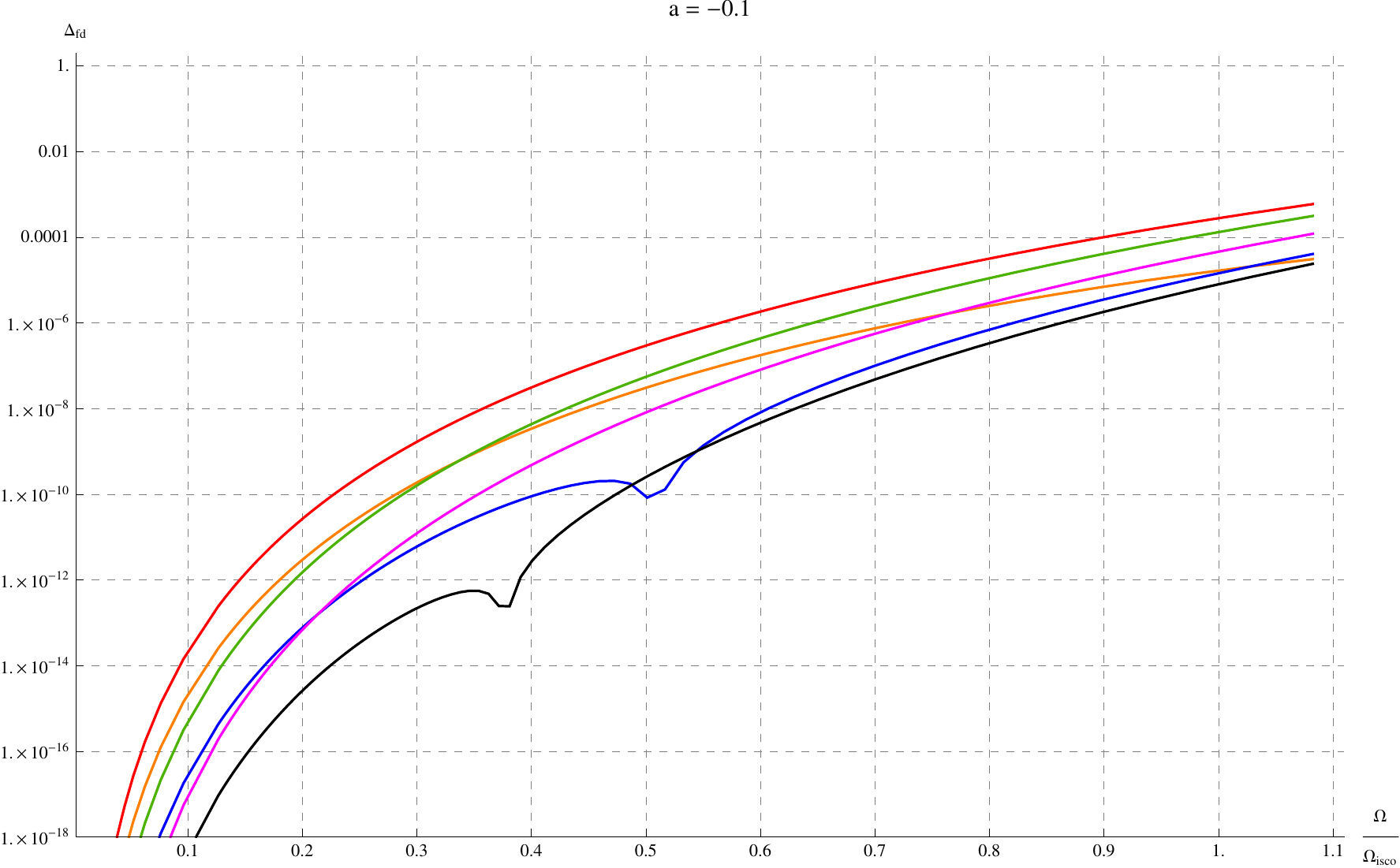}
\par\bigskip\par\bigskip
\includegraphics[width=0.95\textwidth]{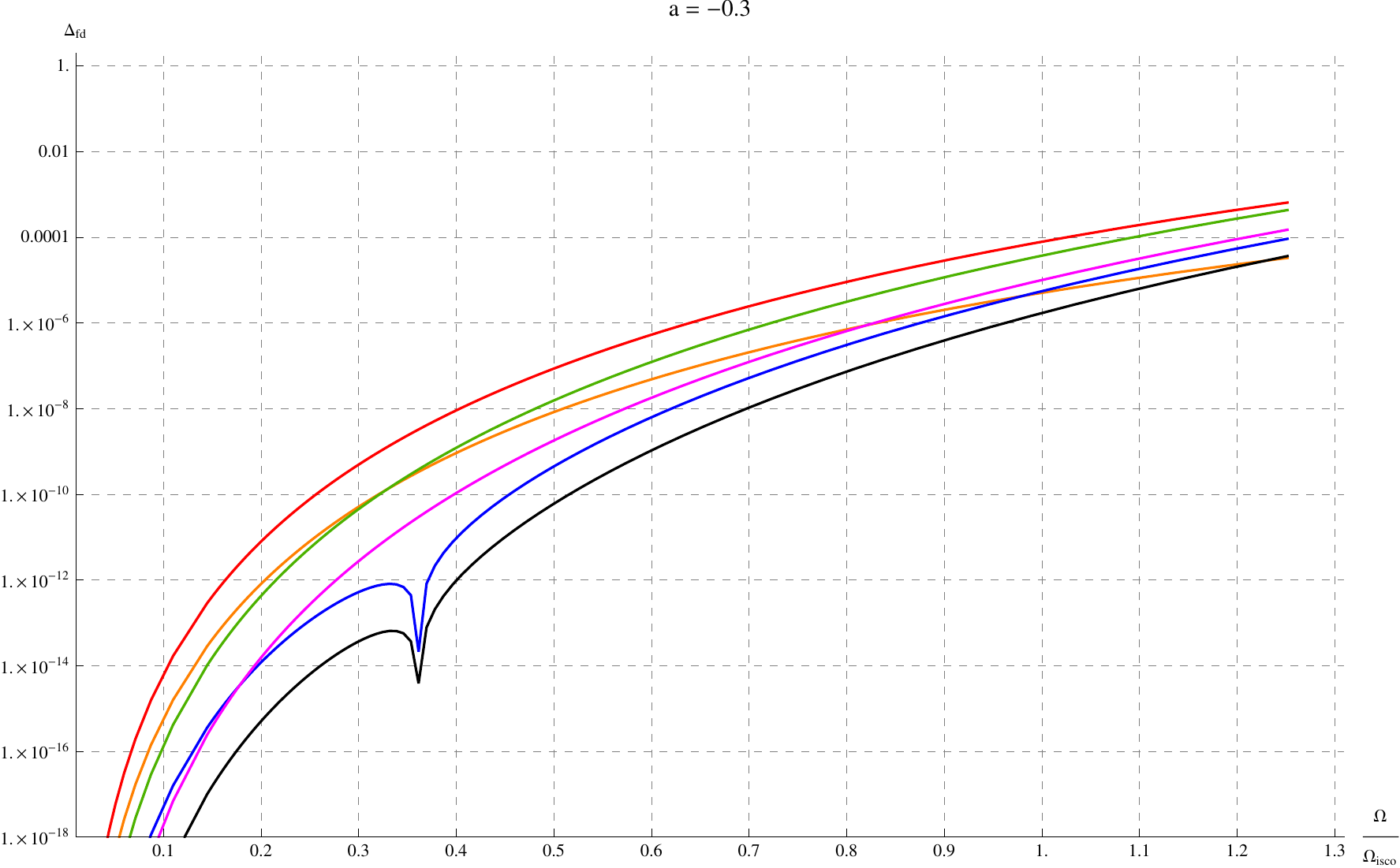}
\caption{Fractional difference ($\Delta_\textrm{fd}$) between the the different pN-order-approximations and the numerical value of the infinity-fluxes at various frequencies compared to the ISCO-frequency for a particular $a$-value. Color-code is as follows - Black: 20pN, Magenta: 19pN, Blue: 18pN, Green: 17pN , Orange: 16pN, Red: 15pN}
\label{fig10}
\end{figure}

\begin{figure}
\centering
\includegraphics[width=0.95\textwidth]{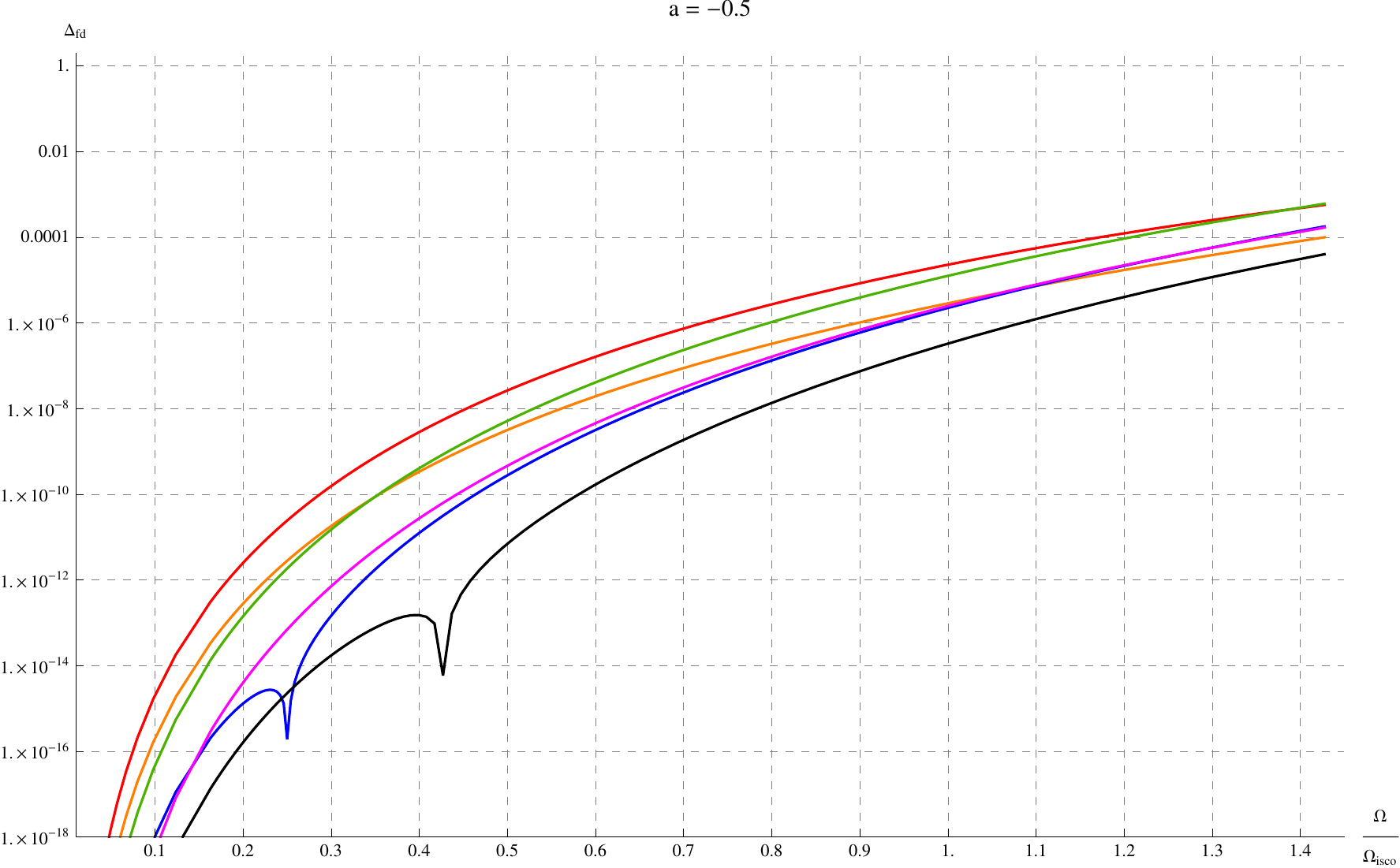}
\par\bigskip\par\bigskip
\includegraphics[width=0.95\textwidth]{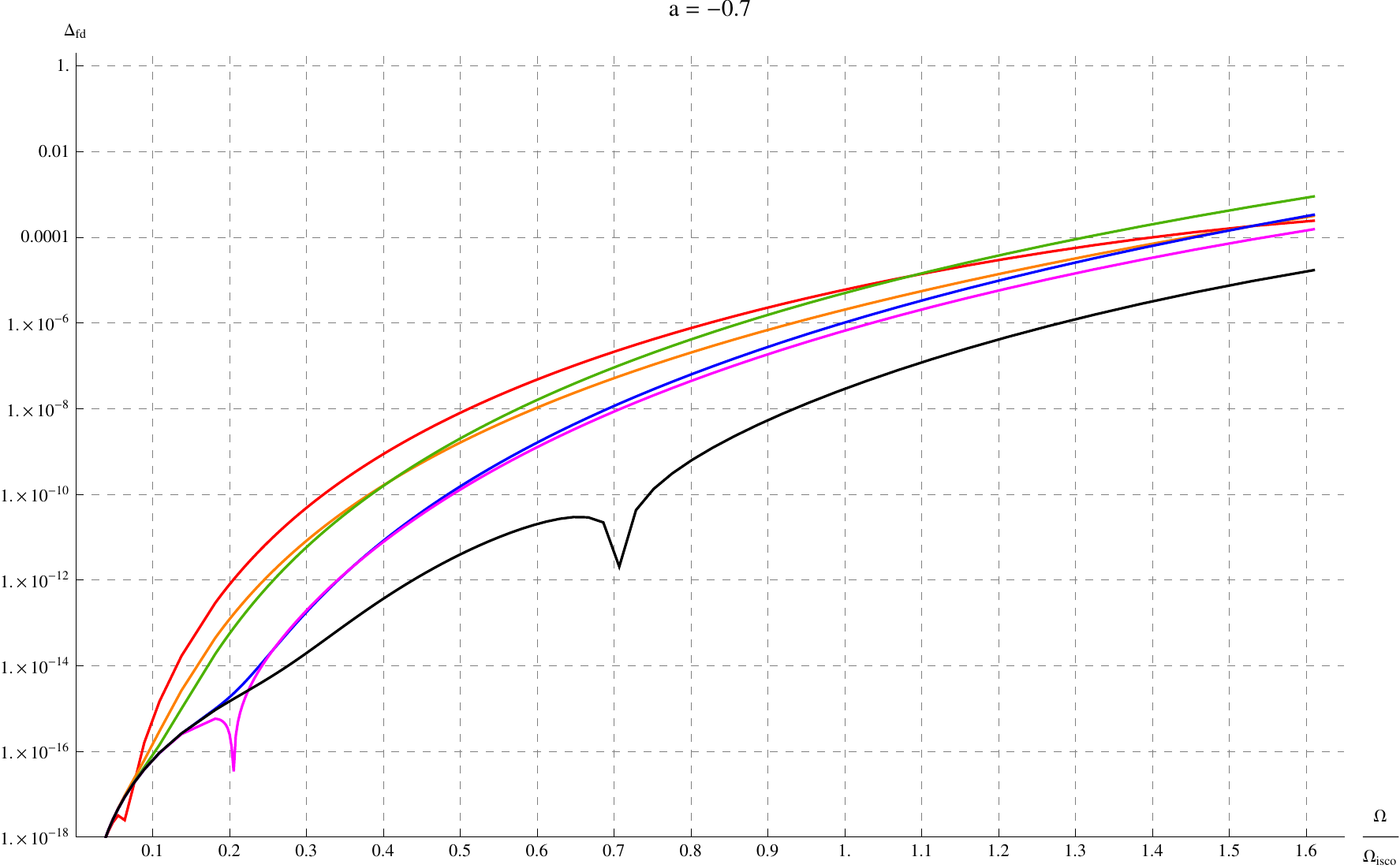}
\caption{Fractional difference ($\Delta_\textrm{fd}$) between the the different pN-order-approximations and the numerical value of the infinity-fluxes at various frequencies compared to the ISCO-frequency for a particular $a$-value. Color-code is as follows - Black: 20pN, Magenta: 19pN, Blue: 18pN, Green: 17pN , Orange: 16pN, Red: 15pN}
\label{fig11}
\end{figure}

\begin{figure}
\centering
\includegraphics[width=0.95\textwidth]{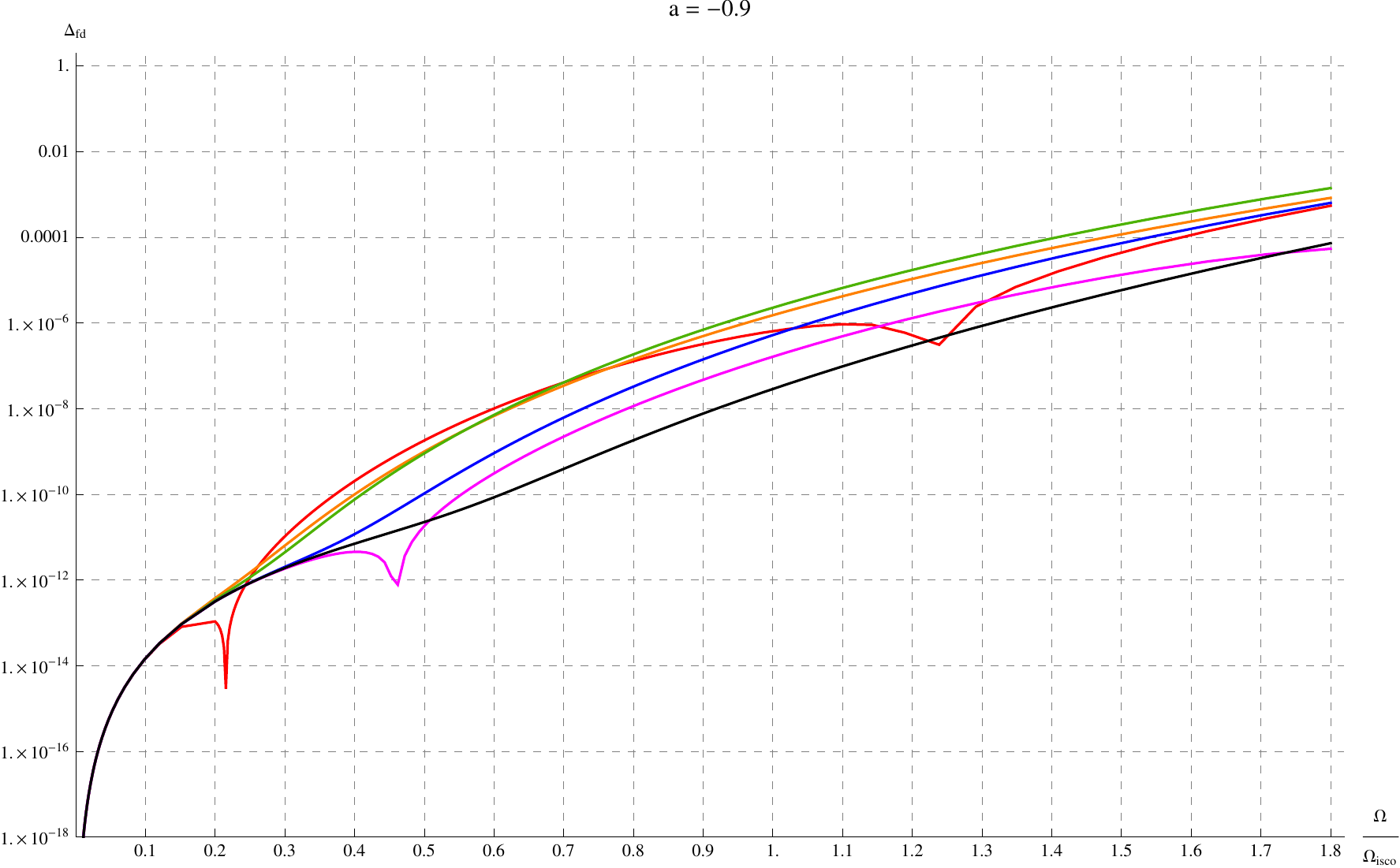}
\par\bigskip\par\bigskip
\includegraphics[width=0.95\textwidth]{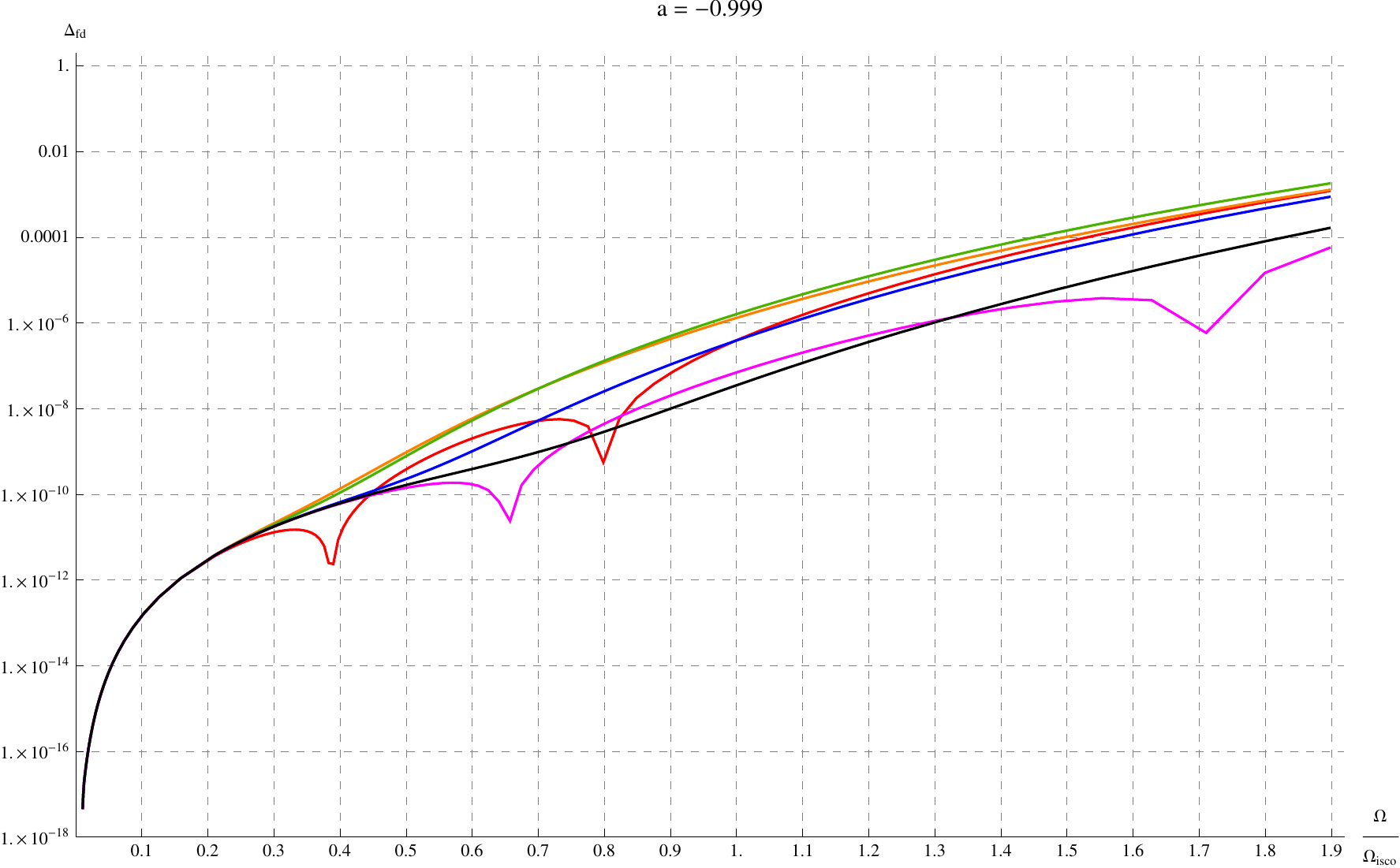}
\caption{Fractional difference ($\Delta_\textrm{fd}$) between the the different pN-order-approximations and the numerical value of the infinity-fluxes at various frequencies compared to the ISCO-frequency for a particular $a$-value. Color-code is as follows - Black: 20pN, Magenta: 19pN, Blue: 18pN, Green: 17pN , Orange: 16pN, Red: 15pN}
\label{fig12}
\end{figure}

In this article, using the analytical solution to the Teukolsky equation in terms of a series over hypergeometric and confluent hypergeometric functions developed by \cite{MST}, we find analytical and numerical pN coefficients of flux at infinity and horizon for a particle in circular, equatorial orbit about a Kerr black hole. Whenever the pN coefficient is not a simple, finite, polynomial in $a$, we write it as a power series in $a$. We find that the power series in $a/M$ for any pN order (up to 18-pN order) is convergent for all values of $a$ with $-1 < a/M < 1$. In Figs (\ref{fig1}$-$\ref{fig12}), we compare the pN-approximated fluxes with their respective numerical values for different spin-values, for both direct and retrograde orbits, at the horizon and infinity. The accuracy with which we compute the fluxes at very large radii stops us from extracting numerical coefficients of higher than 20-pN order. By calculating the fluxes with similar accuracy at lower radii (less than $10^{18}M$) or with higher accuracy at the similar radii, one can extract greater than 20th order pN coefficients. 
%The accuracy of the pN-approximated flux decreases as $R$ decreases (or $\Omega$ increases), and by comparing these pN expressions with numerical results, we find that the 20-pN expressions presented here are accurate for any $R$ up to $1.45R_{ISCO}$ for $a/M=0$, $1.89R_{ISCO}$ for $a/M=0.5$, and $2.97R_{ISCO}$ for $a/M=0.9$ where the accuracy reduces to about 1 part in $10^3$.
%The 20-pN expressions presented here are accurate to $5\%$ for $R=5.3M$,  and the pN approximation gets worse for smaller $R$ or larger $\Omega$.
%At infinity, the 20pN expression is accurate to $5\%$ for $R=4.5M$ and the pN approximation gets worse for smaller $R$ or higher $\Omega$. At the event horizon, the 20pN expression is accurate to $6\%$ for $R=5.1M$. 
%Nagar
Though, in this work we provide the pN expansion of the sum of the multipoles present in the expression for fluxes at infinity and horizon, by \emph{resumming} the individual $(\ell,m)$ modes of these fluxes (which are easily calculable from the numerics that led to this work but not presented here for brevity) like in \cite{DIN,PBFRT}, one can improve the behaviour of the pN expansion in the strong-field limit, allowing one to go to radii below $r_{ISCO}$ and in some cases even closer to the light ring. 
The final goal of this project is to calculate fluxes at infinity and horizon for generic orbits in Kerr spacetime to high pN order and high order in the parameters involved that describe the orbit. To achieve this goal directly would be difficult and we break the path to this goal into three steps. 
The first step involves calculating the flux for slightly eccentric, equatorial orbits about a Kerr black hole  to high pN order where the expansion would be performed in three parameters - the gauge-invariant eccentricity, $\varepsilon$, Kerr spin-parameter, $a$ and the gauge-invariant pN parameter, $R$. Until now this has been calculated to 2.5-pN order and quadratic in $\varepsilon$ in \cite{Kerrecc}. 
The next step would be to calculate the flux for inclined, circular obits about a Kerr black hole to high pN order where the expression would be expanded in three parameters - the angle of inclination, $y$, Kerr spin-parameter, $a$ and the gauge-invariant pN parameter, $R$. Until now, this has been calculated to 2.5-pN order and linear in $y$ in \cite{Kerrinccirc}.
The final step would be to conglomerate the two cases and calculate the flux for eccentric, inclined orbits in Kerr spacetime. 

%{\bf{Note:} } After the completion of this work, we were made aware of Ryuichi Fujita's 10pN order expansion of fluxes at infinity in Kerr (by Soichiro Isoyama).

\label{sec4}

\begin{acknowledgments}
This work was supported by the European Research Council under the European UnionÕs Seventh Framework Programme (FP7/2007-2013)/ERC grant agreement no. 304978 to UoS. The author wishes to thank Leor Barack, John Friedman, Nathan K. Johnson-McDaniel, Maarten van de Meent, Alessandro Nagar and Bernard Whiting for helpful discussions, and is indebted to Nathan K. Johnson-McDaniel for introducing him to previously unknown functions in Mathematica which made analytical extraction of coefficients in this work possible, to Maarten van de Meent for providing his numerical results that led to an initial successful comparison, and to Ryuichi Fujita for sharing his results (\cite{FujitaKerr}) that corrected an earlier typo in the author's work.
\end{acknowledgments}

\appendix
\section{Quantities required to calculate the fluxes}
\label{appA}
To calculate $\alpha$, we need
\bea
&&|C_{\ell,m}|^2 = [ (\lambda+2)^2 + 4am\om-4a^2\om^2 ](\lambda^2+36ma\om-36a^2\om^2) +(2\lambda+3)(96a^2\om^2-48ma\om) + 144\om^2(M^2-a^2), \nonumber \\
&&\varepsilon = \frac{ \sqrt{M^2-a^2} }{ 4Mr_+ }, \nonumber \\
&&k = \om - \frac{m a}{2Mr_+},
\eea
where $\lambda = \mathcal{E}-2am\om+a^2\om^2-2$, and $\mathcal{E}$ is the eigenvalue of the spin-weighted spheroidal harmonic, $_{-2}S_{\ell,m}(\theta,\phi)$ (which is presented by $S$ below). To calculate $\mathcal{E}$ and $S$, we refer the reader to \cite{sf4}.

To calculate $Z^H$ and $Z^\infty$, we need
\begin{align} \allowdisplaybreaks
A_{nn0} &= \frac{-2}{\sqrt{2\pi}\Delta^2}\rho^{-2}\bar{\rho}^{-1}C_{nn}L_1^\dagger [\rho^{-4}L_2^\dagger(\rho^3S)], \nonumber \\
A_{\bar{m}n0} &=  \frac{2}{\sqrt{\pi}\Delta}\rho^{-3}C_{\bar{m}n} \left[  \left(L_2^\dagger S\right)\left( \frac{i K}{\Delta} + \rho + \bar{\rho} \right) - a \sin\theta S\frac{K}{\Delta}(\bar{\rho}-\rho) \right], \nonumber \\
A_{\bar{m}\bar{m}0} &= - \frac{1}{\sqrt{2\pi}} \rho^{-3} \bar{\rho}\, C_{\bar{m}\bar{m}}S \left[ -i \left(\frac{K}{\Delta}\right)_{,r} - \frac{K^2}{\Delta^2} + 2i\rho \frac{K}{\Delta} \right], \nonumber \\
A_{\bar{m}n1} &= \frac{2}{\sqrt{\pi}\Delta} C_{\bar{m}n} \left[  L_2^\dagger S + i a \sin\theta (\bar{\rho}-\rho)S \right], \nonumber \\
A_{\bar{m}\bar{m}1} &= - \frac{2}{\sqrt{2\pi}} \rho^{-3}\bar{\rho}\, C_{\bar{m}\bar{m}}S \left( i \frac{K}{\Delta} + \rho \right), \nonumber \\
A_{\bar{m}\bar{m}2} &= - \frac{1}{\sqrt{2\pi}} \rho^{-3} \bar{\rho}\, C_{\bar{m}\bar{m}}S, \nonumber \\
C_{nn} &= \frac{1}{4\Sigma^3 u^t} \left[ \hat{E}(r^2+a^2) - a \hat{L}_z \right]^2, \nonumber \\
C_{\bar{m}n} &= - \frac{\rho}{2^{3/2}\Sigma^2u^t} \left[ \hat{E}(r^2+a^2) - a \hat{L}_z \right] \left[ i\sin\theta \left( a\hat{E} - \frac{\hat{L}_z}{\sin^2\theta} \right) \right], \nonumber   \displaybreak\\
C_{\bar{m}\bar{m}} &= \frac{\rho^2}{2\Sigma u^t} \left[ i\sin\theta \left( a\hat{E} - \frac{\hat{L}_z}{\sin^2\theta} \right) \right]^2.
\end{align}
where $\Sigma = r^2+a^2\cos^2\theta$, $\rho = (r-ia\cos\theta)^{-1}$, $\Delta = r^2-2Mr+a^2$, $K = (r^2+a^2)\om-am$, and
\bea
L_s^\dagger &=& \partial_\theta - \frac{m}{\sin\theta} + a\omega\sin\theta + s\cot\theta.
\eea
The remaining required quantities are
\bea
&& B^\textrm{trans}_{\ell,m} = \left( \frac{\epsilon\kappa}{\omega} \right)^{-4} e^{i\kappa\epsilon_+\left( 1 + \frac{2\log\kappa}{1+\kappa} \right)} \sum_n a_n \nonumber \\
&& B^\textrm{inc}_{\ell,m} = \omega^{-1} \left( K_\nu - i\epsilon^{-i\pi\nu} \frac{ \sin [\pi(\nu+2+i\epsilon)] }{ \sin [\pi(\nu-2-i\epsilon)] } K_{-\nu-1} \right) A_+ e^{-i\epsilon \left(  \log\epsilon - \frac{1-\kappa}{2}  \right)}\nonumber \\
&& C^\textrm{trans}_{\ell,m} = \omega^{3} A_- e^{i\epsilon \left( \log\epsilon - \frac{1-\kappa}{2} \right)},
\eea
where
\bea
A_+ &=& e^{-\pi\epsilon/2} e^{i\pi(\nu+3)/2}2^{-3-i\epsilon} \frac{ \Gamma(\nu+3+i\epsilon) }{ \Gamma(\nu-1-i\epsilon) } \sum_{n}  a_n, \nonumber \\
A_- &=& e^{-\pi\epsilon/2} e^{-i\pi(\nu-1)/2}2^{1+i\epsilon} \sum_n (-1)^n \frac{ (\nu-1-i\epsilon)_n }{ (\nu+3+i\epsilon)_n } a_n,\,\textrm{and} \nonumber
\eea
\bea
K_\nu = && \frac{ e^{i\epsilon\kappa} (2\epsilon\kappa)^{-2-\nu} 2^{2} \Gamma(3-2i\epsilon_+) \Gamma(2\nu+2) }{ \Gamma(\nu+3+i\epsilon) \Gamma(\nu+1+i\tau) \Gamma(\nu-1+i\epsilon) } \nonumber \\
&\times& \left( \sum_{n=0}^\infty (-1)^n \frac{ \Gamma(n+2\nu+1) \Gamma(n+\nu-1+i\epsilon) \Gamma(n+\nu+1+i\tau) }{ n! \Gamma(n+\nu+3-i\epsilon) \Gamma(n+\nu+1-i\tau) a_n } \right) \nonumber \\
&\times& \left( \sum_{n=-\infty}^0 \frac{ (-1)^n (\nu-1-i\epsilon)_n }{ (-n)! (2\nu+2)_n (\nu+3+i\epsilon)_n } a_n \right)^{-1}.
\eea

\bibliography{sf6}

%%%%%%%%%%%%%%%%%%%%%%%%%%%%%%%%%%%%%%%%%%%%%%%%%%%%%%%%%%%%%%%%%%%%%%%%%%%%%%%%%%%%%%%%%%%%%%%%%%%%%%%%%%%%%%%%%%%%%%%%%%%%%%%%%%%%%%%%%%%%%%%%%%%%%%%%%%%%%%%%%%%%%%%%%%%%%%%%%%%%%%%%%%%%%%%%%%%%%%%%%%%%%%%%%%%%%%%%%%%%%%%%%%%%%%%%%%%%%%%%%%%%%%%%%%%%%%%%%%%%%

\end{document}